\documentclass[prd,aps,twocolumn,showpacs,nofootinbib]{revtex4-1}

\usepackage{amsfonts,color,graphicx,epsfig,amsmath,multirow}
\usepackage[colorlinks=true, linkcolor=blue, citecolor=blue, urlcolor=blue]{hyperref}
\usepackage{mathtools}

\begin{document}

\title{The $B\to K^*$ Transition Form Factors Within the QCD Light-Cone Sum Rules and Their Application to the Semi-leptonic Decay $B \to K^* \mu^+ \mu^-$}

\author{Hai-Bing Fu$^{1,2}$}
\author{Xing-Gang Wu$^{2}$}
\email{wuxg@cqu.edu.cn}
\author{Yang Ma$^{2}$}
\author{Wei Cheng$^{2}$}
\author{Tao Zhong$^{3}$}

\address{$^1$ School of Science, Guizhou Minzu University, Guiyang 550025, P.R. China}
\address{$^2$ Department of Physics, Chongqing University, Chongqing 401331, P.R. China}
\address{$^3$ Department of Physics, Henan Normal University, Xinxiang 453007, P.R. China}

\date{\today}

\begin{abstract}
We present a detailed calculation on the $B\to K^*$ transition form factors (TFFs), $A_{0,1,2}$, $V$ and $T_{1,2,3}$, within the QCD light-cone sum rules (LCSR). To suppress the contributions from high-twist light-cone distribution amplitudes, we adopt a right-handed chiral correlator to do the LCSR calculation. In the resultant LCSRs for the TFFs, the transverse leading-twist distribution amplitude $\phi_{2;K^*}^\bot$ provides over $90\%$ contribution, thus those TFFs provide good platforms for testing the property of $\phi_{2;K^*}^\bot$. We suggest a model for $\phi_{2;K^*}^\bot$, in which two parameters $B_{2;K^*}^\bot$ and $C_{2;K^*}^\bot$ dominantly control its longitudinal distribution. With a proper choice of $B_{2;K^*}^\bot$ and $C_{2;K^*}^\bot$, our predictions on $B\to K^*$ TFFs are consistent with those of lattice QCD predictions. As an application, we also calculate the branching fraction of the $B$-meson rare decay $B \to K^* \mu^+ \mu^-$. The predicted differential branching fraction $d{\cal B}/dq^2(B\to K^{*}\mu^+\mu^-)$ is consistent with the LHCb and Belle measurements within errors. After integrating over the allowable $q^2$-region, we get the branching fraction, ${\cal B}(B\to K^*\mu^+\mu^-) = \left(1.088^{+0.261}_{-0.205} \right)\times 10^{-6}$, where the errors are squared average of the mentioned error sources. When the precision of experimental measurements or the other source of theoretical uncertainties for this process have been further improved in the future, we may get a definite conclusion on the behavior of $\phi_{2;K^*}^\bot$.
\end{abstract}


\maketitle

\section{Introduction}

The $B$-meson rare decays mediated by the penguin-induced flavor-changing-neutral-current transition provide excellent platform for precision test of the standard model (SM) and for probing new physics beyond the SM. Those decays are always suppressed by the loop effects and the Cabibbo-Kobayashi-Maskawa matrix elements, thus, contributions from new interactions could be significant and their effects to the decay rates could be observable.

Among those rare decay channels, the $B$-meson semi-leptonic decay, $B\to K^* \mu^+\mu^-$ with $K^{*} \to K\pi$, arouses people's great interest. Its measurable quantities include the muon forward-backward asymmetry, the longitudinal polarization fraction, the differential and total branching fractions, and etc.. Theoretically, those observables have been studied by using the effective Hamiltonian with or without new physics contributions to the Wilson coefficients, cf.Refs.\cite{Descotes-Genon:2013wba,Bobeth:2010wg, Alok:2010zd, Alok:2011gv, DescotesGenon:2011yn, Bobeth:2011gi, Becirevic:2011bp, Altmannshofer:2011gn, Bobeth:2011nj,Matias:2012xw, Beaujean:2012uj, Altmannshofer:2008dz, Altmannshofer:2012az, Becirevic:2012dx, DescotesGenon:2012zf, Bobeth:2012vn, Descotes-Genon:2013vna}. Experimentally, many measurements have been done by the CDF collaboration~\cite{Aaltonen:2011ja,Aaltonen:2011qs}, the BABAR collaboration~\cite{Lees:2012tva}, the Belle collaboration~\cite{Wei:2009zv}, the LHCb collaboration~\cite{Aaij:2013qta, Aaij:2012cq, LHCb:2012aja, Aaij:2013iag, Aaij:2014bsa,Aaij:2014pli}, the ATLAS collaboration~\cite{ATLAS:2013ola}, and the CMS collaboration~\cite{CMS:cwa}. More specifically, the Belle collaboration gives the total branching fraction ${\cal B}(B\to K^*\ell^+\ell^-) =(1.07^{+0.11}_{-0.10} \pm 0.09)\times 10^{-6}$ ($\ell=e,\mu$)~\cite{Wei:2009zv} and the LHCb collaboration gives ${\cal B}(B\to K^*\mu^+\mu^-) =(1.16\pm0.19) \times 10^{-6}$~\cite{Aaij:2012cq,Aaij:2013qta, LHCb:2012aja}.

Before introducing any new physics, it is better to have a more precise SM prediction. To achieve a precise SM determination on the decay $B\to K^*\mu^+\mu^-$, we are facing the problem of determining the non-perturbative hadronic matrix elements, or equivalently, the $B\to K^*$ transition form factors (TFFs). There are totally seven TFFs for the $B\to$ vector meson decays, which are main error sources for the SM predictions. Table~\ref{tab:formfact} shows the relationships among the non-perturbative matrix elements and the TFFs. As shall be shown latter, $V$ is the TFF defined via a vector current, $A_{0,1,2}$ are the TFFs defined via an axial-vector current, $T_{1}$ is the TFF defined via a tensor current, and $T_{2,3}$ are the TFFs defined via an axial-tensor current. In the literature, those $B\to K^*$ TFFs have been studied under various frameworks, such as the relativistic quark model~\cite{Faessler:2002ut, Ebert:2010dv}, the QCD light-cone sum rules (LCSR)~\cite{Ball:1998kk, Ball:2004rg, Khodjamirian:2006st, Khodjamirian:2010vf, Ball:2005vx, Ali:1999mm}, or the lattice QCD~\cite{Becirevic:2006nm, Liu:2011raa, Horgan:2013hoa}. In the literature, it has also been suggested that one can rearrange those TFFs into more convenient helicity forms, simplifying the LCSR calculations~\cite{Bharucha:2010im,Boyd:1997qw}.

\begin{table}[htb]
\begin{center}
\begin{tabular}{ccl}
\hline
~~Matrix element~~ & ~~TFFs~~ & ~~Relevant decay(s)~~ \\
\hline
$\begin{array}{c}\langle V|\bar{q}\gamma^\mu b|B\rangle
\\ \langle V|\bar{q}\gamma^\mu\gamma^5 b|B\rangle\end{array}$ &
$\begin{array}{c}V\\ A_0, A_1, A_2\end{array}$ &
$\left\}\begin{array}{l} B\to(\rho/\omega)\ell\nu_{\ell} \\
B\to K^*\ell^+\ell^-\end{array}\right.$ \\[5mm]
$\begin{array}{c}\langle V|\bar{q}\sigma^{\mu\nu}q_\nu b|B\rangle \\
\langle V|\bar{q}\sigma^{\mu\nu}\gamma^5 q_\nu b|B\rangle\end{array}$ &
$\begin{array}{c}T_1\\T_2, T_3\end{array}$ &
$\left\}\begin{array}{l} B\to K^*\gamma \\
B\to K^*\ell^+\ell^-\end{array}\right.$ \\ \hline
\end{tabular}
\end{center}
\vspace{-2ex}
\caption{The relations among the $B\to$ vector meson TFFs to the matrix elements and their typical applications for the $B$-meson semileptonic and radiative decays, where $\ell$ stands for the light lepton $e$ or $\mu$.}  \label{tab:formfact}
\end{table}

The QCD LCSR is applicable to low and intermediate $q^2$-region, which can be further extrapolated to all allowable $q^2$-region. The LCSR is based on the operator product expansion (OPE) near the light cone, its non-perturbative dynamics can be parameterized into the light-cone distribution amplitudes (LCDAs) with increasing twists. A more precise LCSR prediction with less theoretical uncertainties shall thus be helpful for a better understanding of those TFFs and the $B\to K^*\mu^+\mu^-$ decay. Several approaches have been adopted to study the vector meson LCDAs, such as the $\rho$- and $K^*$-meson LCDAs~\cite{Ball:1998kk, Huang:2008zg, Choi:2007yu, Ahmady:2014sva, Fu:2014pba, Fu:2014cna}. The vector meson LCDAs have much complex structures, and it is convenient to arrange them via the parameter $\delta\simeq m_{K^*}/m_b\sim0.17$. The relative importance of those LCDAs to the TFFs can be counted by different $\delta$-powers. More specifically, at $\delta^0$-order, we have $\phi^\bot_{2;K^*}$; at $\delta^1$-order, we have $\phi_{2;K^*}^\|$, $\phi_{3;K^*}^\bot$, $\psi_{3;K^*}^\bot$, $\Phi_{3;K^*}^\|$, and $\widetilde\Phi_{3;K^*}^\|$; at $\delta^2$-order, we have $\phi_{3;K^*}^\|$, $\psi_{3;K^*}^\|$, $\Phi_{3;K^*}^\bot$, $\phi_{4;K^*}^\bot$, $\psi_{4;K^*}^\bot$, $\Psi_{4;K^*}^\bot$, and $\widetilde{\Psi} _{4;K^*}^\bot$; at $\delta^3$-order, we have $\phi_{4;K^*}^\|$ and $\psi_{4;K^*}^\|$. Here, the subscripts $2,3,4$ stand for the twist-2, the twist-3 and the twist-4 LCDAs, respectively. For convenience, we present those LCDAs following their $\delta$-powers in Table \ref{DA_delta}. At present, all the $K^*$-meson LCDAs have not been confirmed, especially for the twist-3 and twist-4 ones. Those high-twist LCDAs are suppressed by $\delta^1$-order or higher, however they may provide sizable contribution to the LCSR. Thus, it is helpful to find a proper way to suppress those uncertain sources as much as possible so as to achieve a more precise LCSR prediction.

\begin{widetext}
\begin{center}
\begin{table}[tb]
\begin{tabular}{ cc c  c  }
\hline
  & ~~twist-2~~  & ~~twist-3~~ & ~~twist-4~~  \\
\hline
~~$\delta^0$~~   & $\phi_{2;K^*}^\bot$  &  / & / \\

$\delta^1$         & $\phi_{2;K^*}^\|$ & $\phi_{3;K^*}^\bot$, $\psi_{3;K^*}^\bot$, $\Phi_{3;K^*}^\|$, $\widetilde\Phi_{3;K^*}^\|$  &  /  \\

$\delta^2$         & / & $\phi_{3;K^*}^\|$, $\psi_{3;K^*}^\|$, $\Phi_{3;K^*}^\bot$ & $\phi_{4;K^*}^\bot$, $\psi_{4;K^*}^\bot$, $\Psi_{4;K^*}^\bot$, $\widetilde{\Psi} _{4;K^*}^\bot$ \\

$\delta^3$         &  / & / & $\phi_{4;K^*}^\|$, $\psi_{4;K^*}^\|$ \\
\hline
\end{tabular}
\caption{Following the idea of Ref.\cite{Ball:2004rg}, we write down the $K^*$-meson LCDAs with different twist-structures up to $\delta^3$-order, where $\delta \simeq m_{K^*}/m_b$.} \label{DA_delta}
\end{table}
\end{center}
\end{widetext}

In this paper, by using a proper choice of LCSR correlator, we shall show that the contributions from the higher-twist LCDAs can be highly suppressed or eliminated; especially, all the contributions from the $\delta^1$-order LCDAs are exactly eliminated. Thus, the precision of the QCD LCSRs for $B\to K^*$ TFFs can be greatly improved. Those LCSRs can be inversely adopted as a good platform for determining the properties of $\phi^\bot_{2;K^*}$; e.g. by using those LCSRs, we can fix $\phi^\bot_{2;K^*}$ to a certain degree via a comparison with the lattice QCD predictions and/or a comparison of the experimental data on the $B\to K^*\mu^+\mu^-$ decay.

The remaining parts of the paper are organized as follows. In Sec.II, we present the calculation technology for the $B\to K^*$ TFFs. In Sec.III, we present our numerical results for the $B\to K^*$ TFFs and the differential branching fraction $d{\cal B}/dq^2$ for the $B\to K^*\mu^+\mu^-$ decay. Sec.IV is reserved for a summary. 

\section{Calculation technology}

The $B\to K^*$ TFFs, $A_{0,1,2}$, $V$, $T_{1,2,3}$ and $\widetilde T_3$ can be defined via the follow way,
\begin{widetext}
\begin{eqnarray}
 \langle K^*(p,\lambda )|&&\bar s{\gamma _\mu }(1 - {\gamma _5})b|B(p+q)\rangle = - ie_\mu ^{*(\lambda )}(m_B + m_{K^*} )A_1(q^2) + i(e^{*(\lambda )}\cdot q)\frac{(2p + q)_\mu}{m_B + m_{K^*}}A_2(q^2) \nonumber\\
&&+ iq_\mu (e^{*(\lambda )}\cdot q)\frac{2 m_{K^*} }{q^2}\big[A_3(q^2)- A_0(q^2)\big]  +\epsilon_{\mu\nu\alpha\beta}e^{*(\lambda )\nu} q^\alpha p^\beta \frac{2V(q^2)}{m_B + m_{K^*}},  \label{BKstar:matrix1}
\end{eqnarray}
\begin{eqnarray}
\langle K^*(p,\lambda )|&&\bar s\sigma_{\mu \nu }q^\nu (1 + \gamma_5)b|B(p + q)\rangle = 2i\epsilon_{\mu\nu\alpha\beta}e^{*(\lambda )\nu} q^\alpha p^\beta T_1(q^2) + e_\mu ^{*(\lambda )} (m_B^2 - m_{K^*}^2)T_2(q^2) \nonumber\\
&& - (2p + q)_\mu (e^{*(\lambda )} \cdot q){\widetilde T_3}(q^2)  + q_\mu (e^{*(\lambda )} \cdot q)T_3(q^2),\label{BKstar:matrix2}
\end{eqnarray}
\end{widetext}
where $e^{(\lambda)}$ stands for the $K^*$-meson polarization vector with $\lambda$ being its transverse ($\bot$) or longitudinal ($\|$) component, respectively. $p$ is the $K^*$-meson momentum and $q = p_B - p$ is the momentum transfer between the $B$-meson and the $K^*$-meson. There are some relations among those TFFs, thus not all of them are independent, i.e.
\begin{eqnarray}
A_3(q^2) = \frac{m_B+m_{K^*}}{2 m_{K^*}} A_1(q^2) - \frac{m_B-m_{K^*}}{2m_{K^*}} A_2(q^2),\label{A3}
\end{eqnarray}
\begin{eqnarray}
T_3(q^2) = \frac{m_B^2 - m_{K^*}^2}{q^2} [\widetilde T_3(q^2) - T_2(q^2)]. \label{T3}
\end{eqnarray}
And at the large recoil point $q^2=0$, we have
\begin{eqnarray}
&&A_0(0)=A_3(0) \\
&&T_1(0)=T_2(0)=\widetilde T_3(0)=T(0)\,.\label{T0}
\end{eqnarray}
To derive the LCSRs for those TFFs, we need to deal with the following two correlators:
\begin{widetext}
\begin{eqnarray}
\Pi _\mu^{\rm I} (p,q)=&&i\int d^4x e^{iq\cdot x}\langle{K^*} (p,\lambda)| T\big\{\bar s(x)\gamma_\mu(1-\gamma_5) b(x), j_B^\dag (0)\big\} |0\rangle ,
\label{correlator:1}
\end{eqnarray}
\begin{eqnarray}
\Pi_\mu^{\rm II} (p,q) =&&  - i\int d^4 x e^{i q\cdot x}\langle K^* (p,\lambda )| T\big\{ \bar s(x)\sigma_{\mu\nu} q^\nu (1 + \gamma_5) b(x), j_B^\dag (0) \big\}|0\rangle .
\label{correlator:2}
\end{eqnarray}
\end{widetext}
A natural choice of the current $j_B^\dag (x)$ is $i m_b \bar b(x) \gamma_5 q(x)$, which has the same quantum state as that of the pseudoscalar $B$-meson with $J^{P}=0^-$. Such a choice of correlator shall result in a complex series of all the possible $K^*$-meson twist-structures. For example, up to $\delta^2$-order, all the LCDAs listed in Table \ref{DA_delta} should be taken into consideration in the final LCSRs~\cite{Ball:2004rg}. At present, the properties of the $K^*$-meson LCDAs have not been confirmed, especially for the high-twist LCDAs; it is thus helpful to find a proper way to suppress those uncertain sources as much as possible so as to achieve a more precise LCSR prediction. The LCSR derived with the help of a chiral current~\cite{Huang:1998gp, Huang:2001xb} can be adopted for such purpose. That is, one can choose $j_B^\dag (x)$ as a chiral current, either $i m_b \bar b(x)(1- \gamma_5)q(x)$ or $i m_b \bar b(x)(1+ \gamma_5)q(x)$, to do the calculation. The advantage of such a choice lies in that one can highlight the contributions from different twists of $K^*$-meson LCDAs to the TFFs by selecting a proper chiral current.

In the present paper, we shall adopt the chiral current $j_B^\dag (x)= i m_b \bar b(x)(1 + \gamma_5)q(x)$ to deal with the $B\to K^*$ TFFs. It is noted that the hadronic representation of this correlator contains not only the usual resonance with $J^P=0^-$ but also the extra one with $J^P = 0^+$. This is the price of introducing a chiral correlator for LCSR. But it is worthwhile, since we can eliminate large uncertainty from the twist-2 and twist-3 structures at the $\delta^1$-order and we may also highly suppress the pollution from the scalar resonances with $J^P = 0^+$ by a proper choice of continuum threshold $s_0$. More over, the $J^P = 0^+$ scalar state's contribution, together with the quark-hadron duality approximation, can be further suppressed by applying the Borel transformation. We shall show numerically that the final LCSRs have slight $s_0$ dependence, e.g. if varying $s_0$ from its central value by $\pm 0.5{\rm GeV}^2$, all the TFFs shall be varied by less than five percent. Thus, in the present paper, we shall not discuss the contributions caused by the $0^+$ scalar state.

The correlators, Eqs.(\ref{correlator:1},\ref{correlator:2}), are analytic $q^2$-functions defined at both the time-like and the space-like $q^2$-region. On the one hand, in the time-like $q^2$-region, the long-distance quark-gluon interactions become important and, eventually, the quarks form hadrons. In this region, one can insert a complete series of intermediate hadronic states in the correlator and obtain its hadronic representation by isolating the pole term of the lowest pseudoscalar $B$-meson:
\begin{widetext}
\begin{eqnarray}
\hspace{-20pt}\Pi_\mu^{\rm H(I)}(p,q)&=&\Pi_1^{\rm H (I)} e_\mu^{*(\lambda)} + \Pi_2^{\rm H (I)} (e^{*(\lambda)}\cdot q) (2p+q)_\mu  + \Pi _3^{\rm H (I)} (e^{*(\lambda )}\cdot q)q_\mu  + i\Pi_4^{\rm H (I)} \epsilon_\mu^{\nu \alpha \beta } e_\nu^{*(\lambda )} q_\alpha p_\beta \nonumber\\
&=& \frac{\langle K^*|\bar s \gamma_\mu (1 - \gamma_5)b|B\rangle \langle B|\bar bi m_b\gamma_5 q_1|0\rangle}{m_B^2 - (p + q)^2} + \sum\limits_{\rm H} \frac{\langle K^*|\bar s\gamma_\mu (1 - \gamma_5)b|B^{\rm H}\rangle \langle B^{\rm H}|\bar b i m_b(1 + \gamma _5)q_1|0\rangle}{m_{B^{\rm H}}^2 - (p + q)^2},
\end{eqnarray}
\begin{eqnarray}
\hspace{-20pt}\Pi_\mu^{\rm H(II)}(p,q) &=& i\Pi_1^{\rm H(II)}\epsilon_\mu^{\nu\alpha\beta} e_\nu^{*(\lambda )} q_\alpha p_\beta  + \Pi_2^{\rm H(II)} e_\mu^{*(\lambda )} - \Pi_3^{\rm H(II)}(e^{*(\lambda )} \cdot q)(2p + q)_\mu  + \Pi_4^{\rm H(II)}(e^{*(\lambda )} \cdot q)q_\mu
\nonumber\\
&=& \frac{\langle K^*|\bar s \sigma_{\mu\nu }q^\nu (1+\gamma_5)b|B\rangle \langle B|\bar b i m_b \gamma_5 q_1|0\rangle}{m_B^2 - (p + q)^2}+ \sum\limits_{\rm H} \frac{\langle K^*|\bar s \sigma_{\mu\nu}q^\nu (1 + \gamma _5)b|B^{\rm H}\rangle \langle B^{\rm H}|\bar bi m_b(1 + \gamma _5)q_1|0\rangle }{m_{B^{\rm H}}^2 - (p + q)^2},
\end{eqnarray}
\end{widetext}
where $\langle B|\bar b i m_b\gamma_5 q_1|0\rangle=m_B^2 f_B$ with $f_B$ standing for the $B$-meson decay constant. The invariant amplitudes $\Pi_{1,2,3,4}^{\rm H(I)}$ and $\Pi_{1,2,3,4}^{\rm H(II)}$ are
\begin{eqnarray}
\Pi_1^{\rm H(I)}&&[q^2,(p + q)^2] = \frac{m_B^2f_B(m_B + m_{K^*} )}{m_B^2 -(p + q)^2} A_1(q^2) \nonumber\\
&& + \int_{s_0}^\infty  \frac{\rho_1^{\rm H(I)}}{s - (p + q)^2} ds+{\rm subtractions}, \label{Hadronic:A1}
\end{eqnarray}
\begin{eqnarray}
\Pi _2^{\rm H(I)}&&[q^2,(p + q)^2] = \frac{m_B^2 f_B A_2(q^2)}{(m_B + m_{K^*} )[m_B^2 - (p + q)^2]}\nonumber\\
&& + \int_{s_0}^\infty  \frac{\rho_2^{\rm H(I)}} {s - (p + q)^2} ds +{\rm subtractions},\label{Hadronic:A2}
\end{eqnarray}
\begin{eqnarray}
\Pi_3^{\rm H(I)}&&[q^2,(p + q)^2]  = \frac{2 m_B^2 f_B m_{K^*}[A_3(q^2)-A_0(q^2)]}{q^2 [m_B^2 - (p + q)^2]}\nonumber\\
&&+ \int_{s_0}^\infty  \frac{\rho_3^{\rm H(I)}} {s - (p + q)^2} ds +{\rm subtractions} ,\label{Hadronic:A30}
\end{eqnarray}
\begin{eqnarray}
\Pi_4^{\rm H(I)}&&[q^2,(p + q)^2] = \frac{2 m_B^2 f_B V(q^2)}{(m_B + m_{K^*} )[m_B^2 - (p + q)^2]}\nonumber\\
&& + \int_{s_0}^\infty  \frac{\rho_4^{\rm H (I)}}{s - (p + q)^2} ds +{\rm subtractions}.\label{Hadronic:V}
\end{eqnarray}
\begin{eqnarray}
\Pi_{1,3,4}^{\rm H(II)}&&[q^2,(p + q)^2] = \frac{m_B^2 f_B }{m_B^2 -(p + q)^2} T_i(q^2)  \nonumber\\
&& + \int_{s_0}^\infty  \frac{\rho_{1,3,4}^{\rm H(II)}}{s - (p + q)^2} ds+{\rm subtractions}, \label{Hadronic:T134}
\end{eqnarray}
\begin{eqnarray}
\Pi_2^{\rm H(II)}&&[q^2,(p + q)^2] = \frac{m_B^2f_B(m_B^2 - m_{K^*}^2 )}{m_B^2 -(p + q)^2} T_2(q^2) \nonumber\\
&& + \int_{s_0}^\infty  \frac{\rho_2^{\rm H(II)}}{s - (p + q)^2} ds+{\rm subtractions}. \label{Hadronic:T2}
\end{eqnarray}
For convenience, in Eq.(\ref{Hadronic:T134}), we have used the function $T_i(q^2)\; (i=1,3,4)$ to stand for $2T_1(q^2)$, $\widetilde T_3(q^2)$ and $T_3(q^2)$, respectively. Here we have replaced the contributions of the higher resonances and the continuum states by the dispersion integrations. The continuum threshold parameter $s_0$ is set as the value near the squared mass of the lowest scalar $B$-meson. The spectral densities $\rho_{1,2,3,4}^{\rm H(I,II)}$ can be approximated by applying the conventional quark-hadron duality ansatz
\begin{eqnarray}
\rho_{1,2,3,4}^{\rm H(I,II)} = \rho_{1,2,3,4}^{\rm QCD (I,II)} \theta(s-s_0). \label{quark-hadron-duality}
\end{eqnarray}

On the other hand, in the space-like $q^2$-region, the correlator can be calculated by using the QCD OPE. In this region, we have $(p+q)^2-m_b^2\ll 0$ with the momentum transfer $q^2 \sim {\cal O}(1\;{\rm GeV}^2)\ll m^2_b$, which corresponds to small light-cone distance $x^2\rightsquigarrow0$ and ensures the validity of OPE. The full $b$-quark propagator within the background field approach states
\begin{eqnarray}
 \langle 0|&&{\rm T} \{ b(x)\bar b(0)\} |0\rangle  = i\int \frac{d^4 k} {(2\pi )^4} e^{-ik\cdot x} \frac{\not\! k + m_b}{m_b^2 - k^2}\nonumber\\
&& - i g_s \int \frac{d^4 k}{(2\pi )^4} e^{- ik\cdot x} \int_0^1 dv G^{\mu \nu }(vx) \nonumber\\
&&\times \left[ \frac{1}{2}\frac{\not\! k + m_b} {(m_b^2 - k^2)^2} \sigma_{\mu\nu} + \frac{v}{m_b^2 - k^2} x_\mu \gamma_\nu \right], \label{sq1}
\end{eqnarray}
where $G_{\mu\nu}$ is the gluonic field strength and $g_s$ denotes the strong coupling constant. Using this $b$-quark propagator and carrying out the OPE for the correlator, we obtain the QCD expansion of  $\Pi_\mu^{\rm QCD (I,II)}$ with 2-particle and 3-particle Fock states' contributions,
\begin{widetext}
\begin{eqnarray}
\hspace{-30pt}\Pi_\mu^{\rm QCD (I)}&& = m_b\int \frac{d^4x d^4k}{(2\pi)^4} e^{i(q - k)\cdot x} \bigg\{ \frac{1}{m_b^2 - k^2}\bigg\{ 2k^\mu \langle K^*(p,\lambda)|\bar s(x) q_1(0)|0\rangle  - 2i{k^\nu }\langle K^* (p,\lambda)|\bar s(x)\sigma _{\mu \nu} q_1(0)|0\rangle \nonumber\\
&& - \epsilon_{\mu\nu\alpha\beta} k^\nu \langle K^* (p,\lambda)|\bar s(x)\sigma_{\alpha\beta} q_1(0)|0\rangle \bigg\} - \int dv \bigg\{ \frac{k^\nu}{(m_b^2 - k^2)^2} \bigg[ - i\langle K^*(p,\lambda)|\bar s(x) g_s G_{\mu\nu}(vx)q_1(0)|0\rangle \nonumber\\
&&- 2\langle K^*(p,\lambda)|\bar s(x)\sigma_{\mu\alpha} g_s G^{\alpha \nu}(vx) q_1(0)|0\rangle  + 2i\langle K^*(p,\lambda)| \bar s(x)i g_s {\widetilde G}_{\mu\nu}(vx) \gamma_5 q_1(0)|0\rangle \bigg] \nonumber\\
&& + \frac{2v x_\alpha }{m_b^2 - k^2} \left[-\langle K^*(p,\lambda)|\bar s(x)g_s G_{\mu\alpha}(vx) q_1(0)|0\rangle - i\langle K^*(p,\lambda)|\bar s(x)\sigma_{\mu\beta} g_s G^{\alpha\beta}(vx) q_1(0)|0\rangle \right]\bigg\}\bigg\}, \label{correlator2}
\end{eqnarray}
\begin{eqnarray}
&&\Pi_\mu^{\rm QCD(II)} = i m_b^2 \int\frac{d^4x d^4k}{(2\pi)^4} e^{i(q-k)\cdot x} \bigg\{ \frac{q^\nu}{m_b^2 - k^2}\bigg[2\langle K^*(p,\lambda )| \bar s(x)\sigma_{\mu \nu} q_1(0) |0\rangle  - i\epsilon^{\mu\nu\alpha\beta}\langle K^*(p,\lambda )| \bar s(x) \sigma_{\alpha\beta}q_1(0) |0\rangle \bigg]  \nonumber\\
&&\qquad - \int dv \frac{q^\nu}{(m_b^2 - k^2)^2} [\langle K^*(p,\lambda)|\bar s(x)g G^{\alpha\beta}(vx)\sigma_{\mu \nu}\sigma_{\alpha\beta}q_1(0)|0\rangle  + \langle K^*(p,\lambda)|\bar s(x)g G^{\alpha \beta }(vx){\sigma _{\mu \nu }}{\gamma _5}{\sigma _{\alpha \beta }}q_1(0)|0\rangle ]\bigg\},
\end{eqnarray}
\end{widetext}
where $\widetilde G_{\mu \nu}(vx) = \epsilon_{\mu \nu \alpha \beta} G^{\alpha \beta }(vx)/2$. Up to twist-4 accuracy, the non-zero meson-to-vacuum matrix elements with various $\gamma$-structures, i.e. $\Gamma={\bf 1}$, $i\gamma_5$ and $\sigma_{\mu\nu}$, can be expanded as~\cite{Ball:2007zt}:
\begin{widetext}
\begin{eqnarray}
\langle {K^*} (p,\lambda)|\bar s(x)\sigma_{\mu \nu}q_1(0)|0\rangle &=& - i f_{K^*}^\bot \int_0^1 du e^{iu(p\cdot x)}\bigg\{(e_\mu^{*(\lambda )}p_\nu - e_\nu^{*(\lambda )}p_\mu)\bigg[\phi_{2;{K^*}}^\bot(u)+ \frac{m_{K^*}^2 x^2}{16} \phi_{4;{K^*}}^\bot(u)\bigg]  \nonumber\\
&& +(p_\mu x_\nu - p_\nu x_\mu ) \frac{e^{*(\lambda)}\cdot x}{(p\cdot x)^2} m_{K^*}^2 \left[\phi_{3;{K^*}}^\|(u) - \frac{1}{2}\phi_{2;{K^*}}^\bot(u) - \frac{1}{2} \psi_{4;{K^*}}^\bot(u)\right] \nonumber\\
&& +\frac{1}{2}\left(e_\mu^{*(\lambda )}{x_\nu } - e_\nu^{*(\lambda)}x_\mu\right) \frac{m_{K^*}^2}{p\cdot x} \left[ \psi_{4;{K^*}}^\bot(u)-\phi_{2;{K^*}}^\bot(u) \right] \bigg\},\label{DA1}
\end{eqnarray}
\begin{eqnarray}
\langle {K^*} (p,\lambda )|\bar s(x) q_1(0)|0\rangle &=& - \frac{i}{2} f_{K^*}^\bot\left(e^{*(\lambda)}\cdot x\right) m_{K^*}^2\int_0^1 du e^{iu(p\cdot x)} \psi_{3;{K^*} }^\parallel(u),  \label{DA2}\\
\langle{K^*}(p,\lambda)|\bar s(x)\sigma_{\alpha\beta}g_s G^{\mu\nu}(vx)q_1(0)|0\rangle &=& m_{K^*}^2 f_{K^*}^\bot \frac{e^{*(\lambda )}\cdot x}{2(p \cdot x)}\left[p_\mu\left(p_\alpha g_{\beta\nu}^\bot - {p_\beta}g_{\alpha\nu}^\bot\right) - p_\nu\left(p_\alpha g_{\beta\mu}^\bot-p_\beta g_{\alpha \mu }^\bot\right)\right]\nonumber\\
&&\times \Phi_{3;{K^*} }^\bot(v,p\cdot x), \label{DA4}\\
\langle{K^*}(p,\lambda)|\bar s(x) g_s G^{\mu\nu}(vx)q_1(0)|0\rangle &=& -i m_{K^*}^2 f_{K^*}^\bot \left[e_{\bot\mu}^{*(\lambda)}p_\nu - e_{\bot\nu}^{*(\lambda)}p_\mu\right] \Psi_{4;{K^*}}^\bot(v,p\cdot x), \label{DA5}\\
\langle{K^*}(p,\lambda)|\bar s(x) i{g_s}\tilde G_{\mu\nu} (vx)\gamma_5q_1(0)|0\rangle &=& i m_{K^*}^2 f_{K^*}^\bot \left[e_{ \bot \mu}^{*(\lambda )} {p_\nu } - e_{ \bot \nu }^{*(\lambda )}{p_\mu }\right] \tilde \Psi_{4;{K^*} }^\bot(v,p\cdot x), \label{DA3}
\end{eqnarray}
\end{widetext}
where $f_{K^*}^\bot$ represents the ${K^*}$-meson decay constant,
\begin{displaymath}
\langle {K^*}(p,\lambda)|\bar s(0)\sigma_{\mu \nu }q_1(0)|0\rangle =  i f_{K^*}^\bot ( e_\nu^{(\lambda )}p_\mu-e_\mu^{(\lambda )}p_\nu ),
\end{displaymath}
and we have set
\begin{eqnarray}
g_{\mu\nu}^\bot &=& g_{\mu\nu}-\frac{p_\mu x_\nu + p_\nu x_\mu}{p\cdot x}, \nonumber
\end{eqnarray}
\begin{eqnarray}
e_\mu^\lambda &=& \frac{e^\lambda\cdot x}{p\cdot x} \left( p_\mu - \frac{m_{K^*}^2}{2(p\cdot x)} x_\mu\right)+ e_{\bot\mu}^\lambda,\nonumber
\end{eqnarray}
\begin{eqnarray}
K(v,{p\cdot x}) &=& \int{\cal D}\alpha e^{ipx(\alpha_1+v\alpha_3)}K(\underline{\alpha}).\nonumber
\end{eqnarray}
Here ${\cal D}\alpha=d\alpha_1 d\alpha_2 d\alpha_3\delta(1-\alpha_1 -\alpha_2 -\alpha_3)$ and $K(\underline{\alpha})$ stands for the twist-3 or twist-4 DA $\Phi_{3;K^{*}}^\bot(\underline{\alpha})$, $\Psi_{4;K^{*}}^\bot(\underline{\alpha})$ or $\widetilde\Psi_{4;K^{*}}^\bot(\underline{\alpha})$, in which $\underline{\alpha}=\{\alpha_1,\alpha_2,\alpha_3\}$ corresponds to the momentum fractions carried by the antiquark, quark and gluon, respectively.

Then, by equating the correlators within different $q^2$-regions and by applying the conventional Borel transformation, we obtain the required LCSRs for the $B\to K^*$ TFFs, i.e.
\begin{widetext}
\begin{eqnarray}
A_1(q^2)&=& \frac{m_b m_{K^*}^2 f_{K^*}^\bot}{f_B m_B^2(m_B+m_{K^*})} \bigg\{ \int_0^1\frac{du}{u}e^{\frac{m_B^2- s(u)}{M^2}} \bigg\{\frac{\cal C}{u m_{K^*} ^2}\Theta(c(u,s_0))\phi_{2;{K^*}}^\bot(u,\mu) +\Theta(c(u,s_0)) \psi_{3;{K^*}}^\|(u) -\frac{1}{4}
\nonumber\\
&& \times \bigg[ \frac{m_b^2{\cal C}}{u^3M^4}\widetilde{\widetilde\Theta}(c(u,s_0)) + \frac{{\cal C}-2m_b^2}{u^2M^2}\widetilde\Theta(c(u,s_0))- \frac{1}{u}\Theta(c(u,s_0))\bigg]\phi_{4;{K^*}}^\bot(u)- 2\bigg[\frac{\cal C}{u^2M^2}\widetilde\Theta(c(u,s_0)) -\frac{1}{u}
\nonumber\\
&& \times \Theta(c(u,s_0))\bigg]I_L(u) -\bigg[\frac{2m_b^2}{uM^2}\widetilde \Theta(c(u,s_0)) + \Theta(c(u,s_0)) \bigg] H_3(u)\bigg\} + \int{\cal D}\alpha_i\int_0^1{dv}e^{\frac{m_B^2- s(X)}{M^2}}\bigg[\frac{\underline{\cal C}}{2X^3M^2}
 \nonumber\\
&& -\frac{1}{2X^2}\bigg]\Theta(c(X,s_0)) \left[(4v-1)\Psi_{4;{K^*}}^\bot(\underline\alpha) -\widetilde\Psi_{4;{K^*}}^\bot(\underline \alpha)\right]\bigg\},
\label{TFF_A1}
\end{eqnarray}
\begin{eqnarray}
A_2(q^2) &=& \frac{m_b(m_B + m_{K^*} )m_{K^*}^2 f_{K^*}^\bot }{f_B m_B^2} \bigg\{ \int_0^1 \frac{du}{u} e^{\frac{m_B^2 - s(u)}{M^2}} \bigg\{\frac{1}{m_{K^*} ^2} \Theta(c(u,s_0))\phi_{2;{K^*}}^\bot(u,\mu )- \frac{1}{M^2}\widetilde\Theta (c(u,s_0)) \psi_{3;{K^*}}^{\|}(u)
\nonumber\\
&& - \frac{1}{4}\bigg[\frac{m_b^2}{u^2M^4} \widetilde{\widetilde\Theta}(c(u,s_0)) + \frac{1}{uM^2} \widetilde\Theta(c(u,s_0))\bigg] \phi_{4;{K^*}}^\bot(u) + 2\bigg[\frac{{\cal C} - 2m_b^2}{u^2 M^4}\widetilde{\widetilde\Theta}(c(u,s_0)) - \frac{1}{uM^2}\widetilde\Theta (c(u,s_0))\bigg]
\nonumber\\
&&\times I_L(u) - \frac{1}{M^2} \widetilde \Theta (c(u,s_0))H_3(u)\bigg\}  + \int {\cal D} \alpha_i\int_0^1 dv e^{\frac{m_B^2 - s(X)}{M^2}} \frac{1}{2X^2M^2} \Theta(c(X,s_0))\bigg[(4v - 1)\Psi_{4;{K^*}}^\bot (\underline \alpha)
\nonumber\\
&& - \widetilde \Psi_{4;{K^*}}^ \bot (\underline \alpha) + 4v\Phi_{3;{K^*} }^\bot (\underline \alpha)\bigg]\bigg\}, \label{TFF_A2}
\end{eqnarray}
\begin{eqnarray}
V(q^2) &=& \frac{m_b(m_B + m_{K^*})f_{K^*}^\bot}{f_B m_B^2} \int_0^1 \frac{du}{u} e^{\frac{m_B^2- s(u)}{M^2}} \bigg\{ \Theta(c(u,s_0)) \phi_{2;{K^*}}^\bot(u,\mu) - \frac{m_{K^*}^2}{4} \bigg[\frac{m_b^2}{u^2M^4} \widetilde {\widetilde\Theta}(c(u,s_0)) + \frac{1}{uM^2}
 \nonumber\\
&&\times\widetilde\Theta(c(u,s_0))\bigg] \phi_{4;{K^*}}^\bot (u)\bigg\}, \label{TFF_V}
\end{eqnarray}
\begin{eqnarray}
A_3(q^2)&-&A_0(q^2) = \frac{m_b m_{K^*} f_{K^*}^\bot q^2}{2f_B m_B^2} \bigg\{ \int_0^1 \frac{du}{u} e^{\frac{m_B^2- s(u)}{M^2}} \bigg\{-\frac{1}{m_{K^*}^2}\Theta (c(u,s_0)) \phi_{2;K^*}^\bot(u,\mu) - \frac{2-u}{uM^2}\widetilde \Theta (c(u,s_0))
\nonumber\\
&& \times \psi_{3;K^*}^\| (u) + \frac{1}{4}\bigg[\frac{m_b^2}{u^2M^4} \widetilde{\widetilde\Theta}(c(u,s_0))+ \frac{1}{uM^2} \widetilde\Theta(c(u,s_0))\bigg]\phi_{4;K^*}^\bot(u) + \bigg[(4 - 2u)\bigg[\frac{\cal C}{u^3 M^4} \widetilde{\widetilde\Theta}(c(u,s_0))
\nonumber\\
&&- \frac{2}{u^2 M^2} \widetilde\Theta(c(u,s_0))\bigg] + 2\bigg(\frac{2m_b^2}{u^2 M^4}\widetilde{\widetilde\Theta}(c(u,s_0)) + \frac{1}{uM^2} \widetilde\Theta(c(u,s_0))\bigg)\bigg]  I_L(u)- \frac{2-u}{uM^2}\widetilde\Theta(c(u,s_0))H_3(u)\bigg\}
\nonumber\\
&&   - \int {\cal D} {\alpha_i} \int_0^1 dv e^{\frac{m_B^2- s(X)}{M^2}}\frac{1}{2X^2M^2} \Theta(c(X,s_0)) \bigg[(4v-1) \Psi_{4;K^*}^\bot (\underline\alpha)- \widetilde \Psi_{4;K^*}^ \bot(\underline\alpha) + 4v\Phi_{3;K^*}^\bot(\underline\alpha)\bigg]\bigg\},  \label{A3A0}
\end{eqnarray}
\begin{eqnarray}
T_1(q^2) &=& \frac{m_b^2 m_{K^*}^2 f_{K^*}^\bot }{m_B^2f_B} \bigg\{ \int_0^1  \frac{du}{u} e^{\frac{m_B^2- s(u)}{M^2}} \bigg\{ \frac{1}{m_{K^*}^2}\Theta (c(u,s_0))\phi_{2;K^*}^\bot (u,\mu) - \frac{m_b^2}{4u^2M^4}\widetilde{\widetilde \Theta} (c(u,s_0)) \phi_{4;K^*}^\bot(u)-\frac{2}{uM^2}
\nonumber\\
&&  \times \widetilde\Theta(c(u,s_0)) I_L(u) - \frac{1}{M^2} \widetilde\Theta(c(u,s_0)) H_3(u)\bigg\}+ \int{\cal D} \alpha_i \int_0^1 dv  e^{\frac{m_B^2- s(X)}{M^2}} \frac{5}{4X^2M^2}\widetilde\Theta(c(X,s_0))\Psi_{4;K^*}^\bot(\underline \alpha) \bigg\},  \label{TFF_T1}
\end{eqnarray}
\begin{eqnarray}
T_2(q^2) &=& \frac{m_b^2f_{K^*}^\bot m_{K^*}^2}{m_B^2 f_B} \int_0^1 \frac{du}{u}e^{\frac{m_B^2- s(u)}{M^2}} \bigg\{ \frac{1 - {\mathcal H}}{m_{K^*}^2} \Theta (c(u,s_0)) \phi_{2;K^*}^\bot(u,\mu) - \frac{m_b^2} {4u^2M^4}(1 - {\mathcal H}) \widetilde{\widetilde\Theta}(c(u,s_0)) \phi_{4;K^*}^\bot(u)\nonumber\\
&& - \frac{2(1 - {\mathcal H})}{uM^2} \widetilde\Theta(c(u,s_0))I_L(u) - \frac{1}{M^2}\bigg[ 1 + \bigg( \frac{2}{u}-1 \bigg){\mathcal H} \bigg]\widetilde\Theta (c(u,s_0)) H_3(u) \bigg\}, \label{TFF_T2}
\end{eqnarray}
\begin{eqnarray}
\widetilde T_3(q^2) &=& \frac{m_b^2f_{K^*}^\bot m_{K^*}^2}{m_B^2f_B} \int_0^1\frac{du}{u} e^{\frac{m_B^2- s(u)}{M^2}} \bigg\{ \frac{1}{m_{K^*}^2}\Theta(c(u,s_0))\phi _{2;K^*}^\bot (u) - \frac{m_b^2}{4u^2 M^4}  \widetilde{\widetilde\Theta}(c(u,s_0))\phi_{4;K^*}^\bot(u) - 2\bigg[ \frac{1}{uM^2} \nonumber\\
&&\times \widetilde\Theta(c(u,s_0)) + \frac{2q^2}{u^2M^4}\widetilde{\widetilde\Theta}(c(u,s_0))\bigg]I_L(u) -\frac{1}{M^2}\widetilde \Theta (c(u,s_0))H_3(u)\bigg\}, \label{TFF_T3s}
\end{eqnarray}
\begin{eqnarray}
T_3(q^2) &=& \frac{m_b^2f_{K^*}^\bot m_{K^*}^2}{m_B^2f_B} \int_0^1\frac{du}{u} e^{\frac{m_B^2- s(u)}{M^2}} \bigg\{\frac{1}{m_{K^*}^2} \Theta(c(u,s_0))\phi_{2;K^*}^\bot(u,\mu) - \frac{m_b^2}{4u^2M^4}\widetilde{\widetilde\Theta}(c(u,s_0))\phi_{4;K^*}^\bot(u) - \bigg[\frac{2}{uM^2} \nonumber\\
&& \times \widetilde\Theta(c(u,s_0)) + \frac{4}{u^2 M^4}\widetilde{\widetilde \Theta}(c(u,s_0)) (m_B^2 - m_{K^*}^2)\bigg] I_L(u) + \bigg[ \frac{2}{uM^2} - \frac{1}{M^2} \bigg] \widetilde\Theta(c(u,s_0))H_3(u) \bigg\}, \label{TFF_T3}
\end{eqnarray}
\end{widetext}
where ${\mathcal H} =q^2/(m_B^2-m_{K^*}^2)$ and ${\mathcal C}=m_b^2+u^2m_{K^*}^2-q^2$. $s(\varrho)=[m_b^2-\bar \varrho(q^2-\varrho m_{K^*}^2)]/\varrho$ ($\varrho=u,X$) with $\bar \varrho=1-\varrho$, $X=a_1+a_3$. $c(\varrho,s_0)=\varrho s_0- m_b^2 +\bar\varrho q^2 -\varrho\bar\varrho m_{K^*}^2$. $\Theta(c(\varrho,s_0))$ is the usual step function, $\widetilde\Theta(c(\varrho,s_0))$ and $\widetilde{\widetilde\Theta}(c(\varrho,s_0))$ are defined via the integration
\begin{eqnarray}
\int_0^1 &&\frac{du}{u^2 M^2} e^{-s(u)/M^2}\widetilde\Theta(c(u,s_0))f(u)
\nonumber\\&& = \int_{u_0}^1\frac{du}{u^2 M^2} e^{-s(u)/M^2}f(u) + \delta(c(u_0,s_0))\,,
\label{Theta1}
\end{eqnarray}
\begin{eqnarray}
\int_0^1 && \frac{du}{2u^3 M^4} e^{-s(u)/M^2} \widetilde{\widetilde\Theta}(c(u,s_0))f(u)\nonumber\\&&
= \int_{u_0}^1 \frac{du}{2u^3 M^4} e^{-s(u)/M^2}f(u)+\Delta(c(u_0,s_0))\,.\label{Theta2}
\end{eqnarray}
The surface terms $\delta(c(u_0,s_0))$ and $\Delta(c(u_0,s_0))$ for the 2-particle DAs are
\begin{eqnarray}
\delta(c(u,s_0))&=& e^{-s_0/M^2}\frac{f(u_0)}{{\cal C}_0}, \nonumber\\
\Delta(c(u,s_0))&=& e^{-s_0/M^2}\bigg[\frac{1}{2 u_0 M^2}\frac{f(u_0)} {{\cal C}_0} \nonumber\\
&&\left. -\frac{u_0^2}{2 {\cal C}_0} \frac{d}{du}\left( \frac{f(u)}{u{\cal C}} \right) \right|_{u = {u_0}}\bigg], \nonumber
\end{eqnarray}
where ${\mathcal C}_0 = m_b^2 + {u_0^2}m_{K^*} ^2 - {q^2}$ and $u_0$ is the solution of $c(u_0,s_0)=0$ with $0\leq u_0\leq 1$. There are also surface terms for the 3-particle DAs, however their contributions are quite small and can be safely neglected. The simplified functions $I_L(u)$ and $H_3(u)$ are defined as
\begin{eqnarray}
I_L(u) &=& \int_0^u dv \int_0^v dw \big[\phi_{3;{K^*}}^\|(w) -\frac{1}{2} \phi_{2;{K^*}}^\bot(w,\mu)\nonumber\\&&
-\frac{1}{2} \psi_{4;{K^*}}^\bot(w)\big]\,, \nonumber
\\
H_3(u) &=& \int_0^u dv \left[\psi_{4;{K^*}}^{\bot}(v)-\phi_{2;{K^*}}^\bot(v,\mu)\right]\,.
\end{eqnarray}

\section{Numerical results and discussion} \label{Numerical_analysis}

\subsection{Basic input}

In doing the numerical calculations, we take the $K^*$-meson decay constant $f_{K^*}^\bot=0.185(9)~{\rm GeV}$~\cite{Ball:2007zt}, the $b$-quark pole mass $m_b=4.80\pm0.05\;{\rm GeV}$, the $K^*$-meson mass ${m_{K^*}} = 0.892$GeV, the $B$-meson mass $m_B = 5.279$ GeV~\cite{Agashe:2014kda}, and the $B$-meson decay constant $f_B=0.160\pm0.019{\rm GeV}$~\cite{Fu:2014pba}. The factorization scale $\mu$ is set as the typical momentum transfer of $B\to K^*$, i.e. $\mu\simeq (m_{B}^2-m_b^2)^{1/2} \sim 2.2$ GeV, and its error is estimated by taking $\Delta\mu=\pm1.0$ GeV~\cite{Ball:2004rg}. Such a prediction over the scale changes gives us some idea on the magnitude of the uncalculated next-to-leading order contributions, even though it only predicts part of high-order contributions~\cite{Wu:2013ei}.

As shall be shown latter, the dominant contributions to the TFFs, e.g. Eqs.(\ref{TFF_A1}--\ref{TFF_T3}), are from the transverse leading-twist LCDA $\phi_{2;{K^*}}^\bot$. The chiral-even LCDAs $\phi_{2;{K^*}}^\|$, $\phi_{3;{K^*}}^\bot$, $\psi_{3;{K^*}}^\bot$, $\Phi_{3;{K^*}}^\|$ and $\widetilde\Phi_{3;{K^*} }^\bot$, which are at the $\delta^1$-order, provide zero contributions; and all non-zero twist-3 and twist-4 LCDAs, which are at the $\delta^2$-order, can only provide less than $10\%$ contribution to the total LCSRs. Thus, theoretical uncertainties caused by different choices of the high-twist LCDAs are highly suppressed. For clarity, we take those high-twist LCDAs to be the ones suggested in Ref.\cite{Ball:2007zt}. The dominant ${K^*}$-meson transverse leading-twist LCDA $\phi_{2;{K^*}}^\bot$ can be derived from its light-cone wavefunction (LCWF), which are related via the relation
\begin{equation}
\phi_{2;{K^*}}^\bot(x,\mu_0) = \frac{ 2\sqrt{3}}{ \widetilde{f}_{K^*}^\bot}\int_{|{\bf k}_\bot|^2\leq\mu^2_0}\frac{d{\bf k}_\bot}{16\pi^3}\psi_{2;{K^*}}^\bot(x,{\bf k}_\bot), \label{DA_WF}
\end{equation}
where $\widetilde{f}_{K^*}^\bot = f_{K^*}^{\bot}/{\cal C}_{K^*}^\bot $ is the reduced vector decay constant with ${\cal C}_{K^*}^\bot=\sqrt{3}$.

Following the idea of Ref.\cite{Wu:2010zc}, one can separate the $K^*$-meson LCWF into radial part and spin-space part accordingly, and we call it the WH model for short. The radial part $\psi_{2;{K^*}}^{R}$ can be constructed from the Brodsky-Huang-Lepage prescription~\cite{BHL}, and the spin-space part $\chi_{K^*}^{h_1 h_2}(x,\bf{k}_\bot)$ is from the Wigner-Melosh rotation~\cite{Cao:1997hw}. More specifically, the $K^*$ meson WH LCWF states
\begin{eqnarray}
\psi_{2;K^*}^\bot(x,{\bf k}_\bot ) = \sum\limits_{h_1 h_2} {{\chi_{K^*}^{{h _1}{h_2}}}} (x,{{\bf{k}}_ \bot }) \psi _{2;{K^*}}^{R}(x,{{\bf{k}}_ \bot }),
\end{eqnarray}
where
\begin{eqnarray}
\psi_{2;K^*}^R &\propto & [1 + B_{2;K^*}^\bot C_1^{3/2}(\xi)+ C_{2;K^*}^\bot C_2^{3/2}(\xi)]\nonumber\\
&\times & \exp\left[-b_{2;K^*}^{\bot 2} \left(\frac{{\bf k}_\bot^2 + m_s^2}{x}
+ \frac{{\bf k}_\bot^2 + m_q^2}{\bar x} \right) \right],  \label{psiR}
\end{eqnarray}
where $C_{1,2}^{3/2}$ are Gegenbauer polynomials, $\xi=2x-1$, $x$ stands for the $s$-quark momentum fraction of the meson and $\bar{x}=1-x$ stands for that of the light-quark $q$. $m_s$ is the $s$-quark mass and $m_q$ is the light-quark mass. The spin-space wavefunction
\begin{eqnarray}
\chi _{K^*}^{h_1 h_2} (x,{\bf k}_\bot ) = \frac{\bar x m_s + x m_q} {\sqrt {{\bf k}_\bot^2 + (\bar x m_s + x m_q)^2} }.
\end{eqnarray}
Then, one can get the WH-DA
\begin{eqnarray}
&&\phi_{2;K^*}^\bot(x,{\mu _0}) = \frac{A_{2;K^*}^\bot \sqrt {3x\bar x} {\rm Y}} {8 \pi^{3/2} \tilde f_{K^*}^\bot b_{2;K^*}^\bot } [1 + B_{2;K^*}^\bot C_1^{3/2}(\xi )  \nonumber\\
&&~~ +C_{2;K^*}^\bot C_2^{3/2}(\xi )]\exp\left[ - b_{2;K^*}^{\bot 2} \frac{ \bar x m_s^2 + x m_q^2 - {\rm Y}^2 }{x\bar x} \right]  \nonumber\\
&&~~\times\left[ {\rm Erf} \bigg( b_{2;K^*}^\bot \sqrt{\frac{\mu _0^2 + {\rm Y}^2} {x\bar x}}  \bigg) - {\rm Erf}\left( b_{2;K^*}^\bot \sqrt{\frac{{\rm Y}^2}{x\bar x} } \right) \right],\nonumber\\
\label{DA_WH}
\end{eqnarray}
where $\textrm{Erf}(x) =\frac{2}{\sqrt{\pi}}\int^x_0 e^{-t^2} dt$, ${\rm Y}=\bar x m_s + x m_q$ and the constituent quark mass $m_q\simeq300$ MeV and $m_s\simeq450$ MeV. In addition to the normalization condition, the average value of the squared transverse momentum can be regarded as another constraint, which is defined as
\begin{eqnarray}
\langle{\bf k}_\bot^2\rangle_{2;{K^*}}^{1/2}=\frac{\int dx d^2{\bf k}_\bot |{\bf k}_\bot|^2 |\psi_{2;{K^*}}^\bot(x,{\bf k}_\bot)|^2}{\int dx d^2 {\bf k}_\bot|\psi_{2;{K^*}}^\bot(x,{\bf k}_\bot)|^2}.
\end{eqnarray}
Here, the value of $\langle{\bf k}_\bot^2\rangle_{2;{K^*}}^{1/2} \sim 0.37 {\rm GeV}^2$~\cite{Wu:2010zc}.

Conventionally, the light meson's transverse leading-twist LCDA can be expanded as a series of Gegenbauer polynomials. The Gegenbauer moments $a_n^\bot$ at the initial scale $\mu_0$ can be calculated via the formula
\begin{equation}
a_n^\bot(\mu_0^2)=\frac{\int_0^1 dx~ \phi_{2;{K^*}}^\bot(x,\mu_0^2) C_n^{3/2}(\xi)}{\int_0^1 dx~ 6x \bar x[C_n^{3/2}(\xi)]^2},
\end{equation}
where $\mu_0 \sim 1$ GeV stands for some initial scale. To next-to-leading order accuracy, the scale dependence of the Gegenbauer moments $a_n^\bot$ can be written as~\cite{Ball:2006nr}
\begin{eqnarray}
 a^{\bot \rm NLO}_n(\mu^2) &=&  a_n^\bot(\mu_0^2) E_n^{\rm NLO}\nonumber\\
&+&\frac{\alpha_s(\mu^2)}{4\pi}\sum_{k=0}^{n-2} a_k(\mu_0^2)\,
L^{\gamma_k^{(0)}/(2\beta_0)}\, d^{(1)}_{nk}
\end{eqnarray}
where
\begin{eqnarray}
E_n^{\rm NLO} &=&  L^{\gamma^{(0)}_n/(2\beta_0)}\times \nonumber\\
&&\left\{1+ \frac{\gamma^{(1)}_n \beta_0-\gamma_n^{(0)}\beta_1}{8\pi\beta_0^2}
\Big[\alpha_s(\mu^2)-\alpha_s(\mu_0^2)\Big]\right\}
\end{eqnarray}
with $L=\alpha_s(\mu^2)/\alpha_s(\mu^2_0)$, $\beta_0=11-2n_f/3$ and $\beta_1=102-38n_f/3$. $\gamma_n^{(0),(1)}$ are anomalous dimensions~\cite{Ball:2006nr}. The one-loop anomalous dimension $\gamma^\bot_{{K^*}}=8C_F[\psi(n+2)+\gamma_E-3/4]$ with $\psi(n+1)=\sum_{k=1}^n 1/k -\gamma_E$.

\begin{table}[tb]
\begin{center}
\begin{tabular}{c c | c c c c}
\hline
~$a_1^{\bot}$~~ &  ~~$a_2^{\bot}$~ & ~$B_{2;{K^*}}^\bot$~    &  ~$C_{2;{K^*}}^\bot$~ & ~$A_{2;{K^*}}^\bot$~ & ~$b_{2;{K^*}}^\bot$~  \\ \hline
  $0.04$  & $0.10$  &   0.0038  &   0.135   &   30.61   &   0.625   \\
  $0.01$  & $0.10$  &   $-0.027$  &   0.138   &   30.81   &   0.626   \\
  $0.07$  & $0.10$  &   0.036   &   0.132   &   30.39   &   0.623   \\
  $0.04$  & $0.02$  &   0.008   &   0.056   &   32.55   &   0.648   \\
  $0.04$  & $0.18$  &   $-0.003$  &   0.214   &   28.53   &   0.599   \\
\hline
\end{tabular}
\caption{The ${K^*}$-meson transverse leading-twist LCDA parameters $A_{2;{K^*}}^\bot$ and $b_{2;{K^*}}^\bot$ under some typical choices of $a_{1,2}^{\bot}$ at initial scale $\mu_0 = 1{\rm GeV}$.}\label{DA_parameter_1}
\end{center}
\end{table}

\begin{figure}[htb]
\centering
\includegraphics[width=0.4\textwidth]{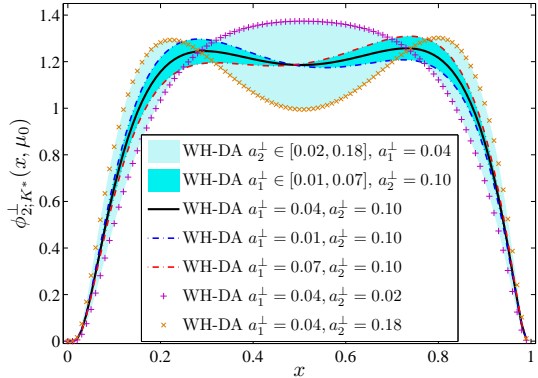}
\caption{The WH-DA model under the condition of $a^\perp_1 \in [0.01, 0.07]$ and $a^\perp_2 \in [0.02, 0.18]$. } \label{DA:1bot}
\end{figure}

The WH-model has four undetermined parameters $A_{2;K^*}^\bot$, $b_{2;K^*}^\bot$, $B_{2;K^*}^\bot$ and $C_{2;K^*}^\bot$. In addition to the normalization condition and the squared transverse momentum $\langle{\bf k}_\bot^2\rangle_{2;{K^*}}^{1/2}$, we shall adopt the values of the first two Gegenbauer moments $a_1^\bot(1{\rm GeV})=0.04(3)$ and $a_2^\bot(1{\rm GeV})=0.10(8)$~\cite{Ball:2007zt} as further criteria. The corresponding values are listed in Table \ref{DA_parameter_1}. Qualitatively, the parameter $B_{2;K^*}^\bot$ dominants the $K^*$-meson ${\rm SU}_f(3)$-breaking effect, a larger $|B_{2;K^*}^\bot|$ indicates a larger breaking effect; the parameter $C_{2;K^*}^\bot$ dominants the shape of the LCDA, a larger $C_{2;K^*}^\bot$ indicates a double-peak behavior and a smaller one tends to a single-peaked behavior.

\begin{figure}[htb]
\centering
\includegraphics[width=0.4\textwidth]{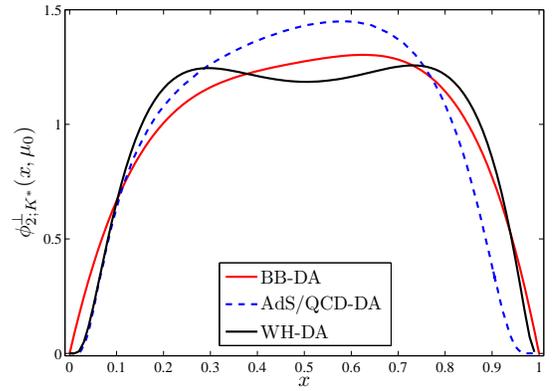}
\caption{A comparison of $\phi_{2;{K^*}}^\bot(x,\mu_0=1{\rm GeV})$ under various models. For the WH-DA model and the BB-DA model, we take $a^\perp_1 =0.04$ and $a^\perp_2 =0.10$. }
\label{DA:2bot}
\end{figure}

In addition to the WH-DA model, there are other $\phi_{2;K^*}^\bot$ models which have also been suggested in the literature. Following the standard Gegenbauer expansion, it has been suggested by Ball and Braun (we call it the BB-DA model)~\cite{Ball:2007zt}, $a_1^\bot(1{\rm GeV})=0.04(3)$ and $a_2^\bot(1{\rm GeV})=0.10(8)$. Another typical model based on the AdS/QCD theory has been suggested in Refs.\cite{deTeramond:2012rt, Brodsky:2013npa}, we call it the AdS/QCD-DA model. We put the twist-2 LCDA $\phi_{2;K^*}^\bot(x,\mu_0)$ in Fig.~\ref{DA:2bot}, in which the WH-DA, BB-DA and AdS/QCD-DA models are presented as a comparison. The small asymmetries of those LCDAs indicate small $K^*$-meson ${\rm SU}_f(3)$-breaking effects. If we have precise measurements for the processes involving $K^*$-meson, the $\phi_{2;K^*}^\bot$ behavior can be determine by comparing theoretical predictions with the data.

\begin{figure}[htb]
\centering
\includegraphics[width=0.45\textwidth]{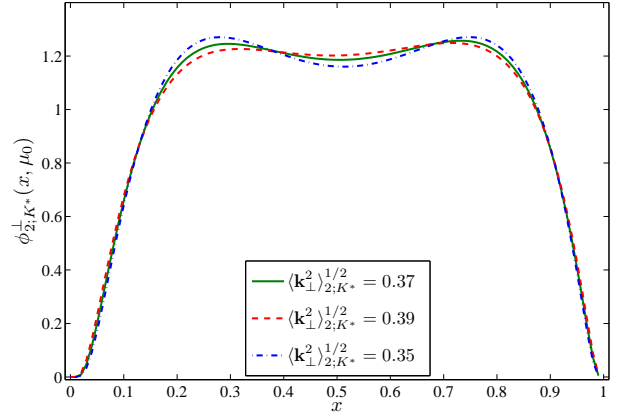}
\caption{The $\phi_{2;{K^*}}^\bot(x,\mu_0=1{\rm GeV})$ by setting $\langle{\bf k}_\bot^2\rangle_{2;{K^*}}^{1/2} = 0.37 \pm 0.02$, where $a^\perp_1 =0.04$ and $a^\perp_2 =0.10$. }
\label{DA:3}
\end{figure}

We make a simple discussion of how the value of $\langle{\bf k}_\bot^2\rangle_{2;{K^*}}^{1/2}$ affects the LCDA behavior. For the purpose, we set $\langle{\bf k}_\bot^2\rangle_{2;{K^*}}^{1/2}=0.37\pm0.02{\rm GeV}^2$. The results are presented in Fig.~\ref{DA:3}. It indicates that the LCDA depends slightly on the choice of $\langle{\bf k}_\bot^2\rangle_{2;{K^*}}^{1/2}$. Such a small effect shall further brings small uncertainty to the TFFs. For example, at large recoil region $q^2=0$, it only brings about less than $\sim1\%$ contributions to the TFFs. Thus in the following, we shall fix $\langle{\bf k}_\bot^2\rangle_{2;{K^*}}^{1/2}=0.37{\rm GeV}^2$.

\subsection{The $B \to K^*$ TFFs at low $q^2$-region}

\begin{table}[htb]
\begin{tabular}{ c c c c c c }
\hline
$s_0^{A_1}$ & $M^2_{A_1}$ & $s_0^{A_2}$ & $M^2_{A_2}$ & $s_0^{A_{3-0}}$ & $M^2_{A_{3-0}}$ \\
34.9(5) & 8.3(5) & 32.5(5) & 7.7(5) & 35.5(5) & 7.5(5)\\
\hline
$s_0^{V}$ & $M^2_V$  & $s_0^{T_1}$ & $M^2_{T_1}$ & $s_0^{T_3}$ & $M^2_{T_3}$\\
35.9(5) & 11.6(5) & 37.2(5) & 13.2(5) & 34.2(5) & 9.7(5)\\
\hline
\end{tabular}
\caption{The determined $B \to K^*$ continuum threshold $s_0$ and the Borel parameter $M^2$ at the large recoil point for the WH-DA model. The central values are for $m_b = 4.80$ GeV.}\label{s0M2}
\end{table}

We adopt the following criteria to set the LCSR parameters, such as the Borel window and $s_0$, for the $B\to {K^*}$ TFFs. First, we require the continuum contribution to be less than $30\%$ of the total LCSR. Second, we require all high-twist DAs' contributions to be less than $15\%$ of the total LCSR. Third, the derivatives of the LCSRs Eqs.(\ref{TFF_A1}--\ref{TFF_T3}) with respect to $(-1/M^2)$ provide the LCSRs for $m_B$. To be self-consistent, we require all the predicted $B$-meson masses from those LCSRs to be full-filled in comparing with the experimental one, e.g. ${|m_B^{\rm LCSR}- m_B^{\rm exp}|}/{m_B^{\rm exp}}\leq 0.1\%$. The determined continuum threshold $s_0$ and the Borel parameter $M^2$ for the $B\to {K^*}$ TFFs at the large recoil point $q^2=0$ are presented in Table~\ref{s0M2}. Numerically we have found that the uncertainties caused by different choices of $s_0$ are small for all the TFFs, which are less than $5\%$. More explicitly, by varying $s_0$ from its central value by $\pm 0.5{\rm GeV}^2$, we obtain $A_1(0)=0.290\pm 0.009~{\rm GeV}$, $A_2(0)=0.257^{+0.014}_{-0.012}~{\rm GeV}$, $V(0) = 0.372\pm 0.011~{\rm GeV}$, $T_1(0)=T_2(0)=\widetilde T_3(0)=0.351^{+0.007}_{-0.009}~{\rm GeV}$, and $T_3=0.236^{+0.010}_{-0.005}~{\rm GeV}$.

\begin{figure}[htb]
\centering
\includegraphics[width=0.45\textwidth]{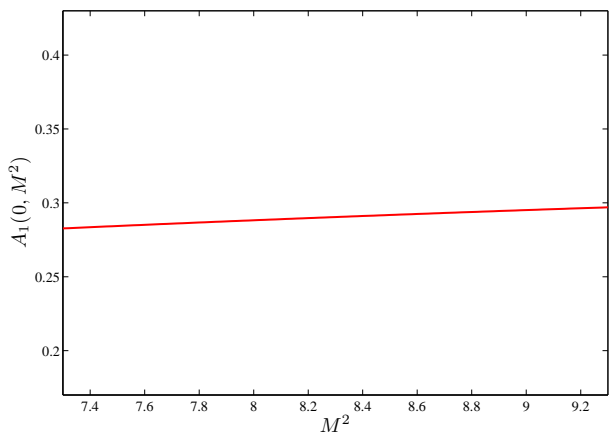}
\caption{The TFF $A_1(0)$ versus the Borel parameter $M^2$.}
\label{A1M2}
\end{figure}

In the literature, the flatness of the TFF is sometimes adopted as a criterion to set the Borel window of the sum rules, cf. a review on the QCD sum rules~\cite{Colangelo:2000dp}. As a reference, we take the TFF $A_1(0)$ as an example to show how the TFF changes with variation of $M^2$. The conditions for other TFFs are similar. We preset the TFF $A_1(0)$ versus $M^2$ in Fig.~\ref{A1M2}. Fig.(\ref{A1M2}) indicates that $A_1(0)$ is almost unchanged within the allowable Borel window $M^2=8.3\pm0.5$, then our present predictions are consistent with the flatness criterion.

\begin{table}[tb]
\begin{tabular}{c c c c c}
\hline
 & WH & BZ~\cite{Ball:2004rg} & LCSR~\cite{Khodjamirian:2010vf} & AdS~\cite{Ahmady:2014sva} \\ \hline

$A_1$    &   $0.290^{+0.029}_{-0.031}$    &   $0.292\pm0.028$    &   $0.25^{+0.16}_{-0.10}$  &   0.249   \\
$A_2$    &   $0.257^{+0.037}_{-0.042}$    &   $0.259\pm0.027$    &   $0.23^{+0.19}_{-0.10}$  &   0.235   \\
$A_0$    &   $0.372^{+0.143}_{-0.141}$    &   $0.374\pm0.034$    &   $0.29^{+0.10}_{-0.07}$  &   0.285   \\
$V$      &   $0.411^{+0.043}_{-0.045}$    &   $0.411\pm0.033$    &   $0.36^{+0.23}_{-0.12}$  &   0.277   \\
$T_1[T_2, \widetilde T_3]$ &   $0.351^{+0.036}_{-0.035}$   &   $0.333\pm0.028$ &   $0.31^{+0.18}_{-0.10}$  &   0.255   \\
$T_3$    &   $0.236^{+0.032}_{-0.033}$    &   $0.202\pm0.018$    &   $0.22^{+0.17}_{-0.10}$  &   0.155   \\
\hline
\end{tabular}
\caption{The $B\to {K^*}$ TFFs at the large recoil point $F_i(0)$, where the errors are squared average of all mentioned error sources. As a comparison, the results derived by Ball and Zwicky (BZ)~\cite{Ball:2004rg}, the AdS/QCD~\cite{Ahmady:2014sva} predictions, and the LCSR predictions~\cite{Khodjamirian:2010vf} are also presented. } \label{F0_Bq_1}
\end{table}

\begin{table}[htb]
\begin{tabular}{ c c c c }
\hline
$\mu$ & $1.2{\rm GeV}$ & $2.2{\rm GeV}$ & $3.2{\rm GeV}$ \\ \hline
$A_1$  & 0.294 	&	0.290 	&	0.290 \\
$A_2$  & 0.264 	&	0.257 	&	0.257 \\
$V$    & 0.416 	&	0.411 	&	0.411 \\
$A_0$  & 0.386 	&	0.372 	&	0.372 \\
$T_1[T_2, \widetilde T_3]$  & 0.355 	&	0.351 	&	0.351 \\
$T_3$  & 0.241 	&	0.236 	&	0.236 \\ \hline
\end{tabular}
\caption{The $B\to {K^*}$ TFFs at large recoil point $F_i(0)$ for different choice of scale, i.e. $\mu=(2.2\pm1.0)$ GeV. } \label{Factorization_scale}
\end{table}

We present the $B\to K^*$ TFFs at large recoil point $q^2=0$ in Table \ref{F0_Bq_1}, where, as a comparison, we also give the predictions by Ball and Zwicky (BZ)~\cite{Ball:2004rg}, the AdS/QCD predictions by Ref.\cite{Ahmady:2014sva} and the LCSR predictions by Ref.\cite{Khodjamirian:2010vf}. For the WH model, the errors are squared averages of all the above mentioned error sources. To have a clear look at the factorization scale dependence, in Table \ref{Factorization_scale}, we list the $B\to K^*$ TFFs at large recoil point under the choice of $\mu=(2.2\pm 1.0)$ GeV. To be consistent, the LCDAs shall be run to different scales via the usual one-loop QCD evolution equation. Table \ref{Factorization_scale} shows that the factorization scale dependence is small (the TFFs are almost unchanged when $\mu>2.2$ GeV), which are less than $4\%$ for all the $B\to K^*$ TFFs.

Table \ref{F0_Bq_1} shows that even though, we have taken different correlators to do the LCSR calculation in comparison to that of Ref.\cite{Ball:2004rg}, the central values of our present LCSRs agree with those of Ref.\cite{Ball:2004rg} under the choice of similar twist-2 LCDA $\phi_{2;{K^*}}^\bot$. This agreement could be treated as a cross check of different LCSR calculations. It is noted that our present theoretical uncertainties are somewhat larger than those of Ref.\cite{Ball:2004rg}. This is reasonable, since by using the fixed higher-twist LCDAs \footnote{A rough discussion on the uncertainties of higher-twist LCDAs has been given in Ref.\cite{Ball:2004rg}.}, the largest errors of those LCSRs are from the leading-twist LCDA $\phi_{2;{K^*}}^\bot$, while our present LCSRs in effect amplify such errors. In fact, the higher-twist LCDAs at $\delta^1$-order are mostly uncertain, and their errors shall potentially provide large uncertainties to the LCSRs derived by using the usual current $i m_b \bar b(x) \gamma_5 q(x)$. On the other hand, by using the chiral correlators, as shown by Eqs.(\ref{correlator:1}, \ref{correlator:2}), those high-twist LCDAs' contributions are greatly suppressed, then the accuracy of the LCSRs can be greatly improved. For example, we find that the contributions from the twist-3 LCDA $\Phi_{3;K^*}^\|$ and the twist-4 LCDA $\Psi_{4;K^*}^\bot$ provide less than $0.1\%$ of the total LCSRs, thus their own uncertainties are highly suppressed. Because of the suppression of the most uncertain high-twist contributions, our present LCSRs provide a good platform for determining the accurate behavior of $\phi_{2;{K^*}}^\bot$.

\begin{figure*}
\begin{center}
\includegraphics[width=0.24\textwidth]{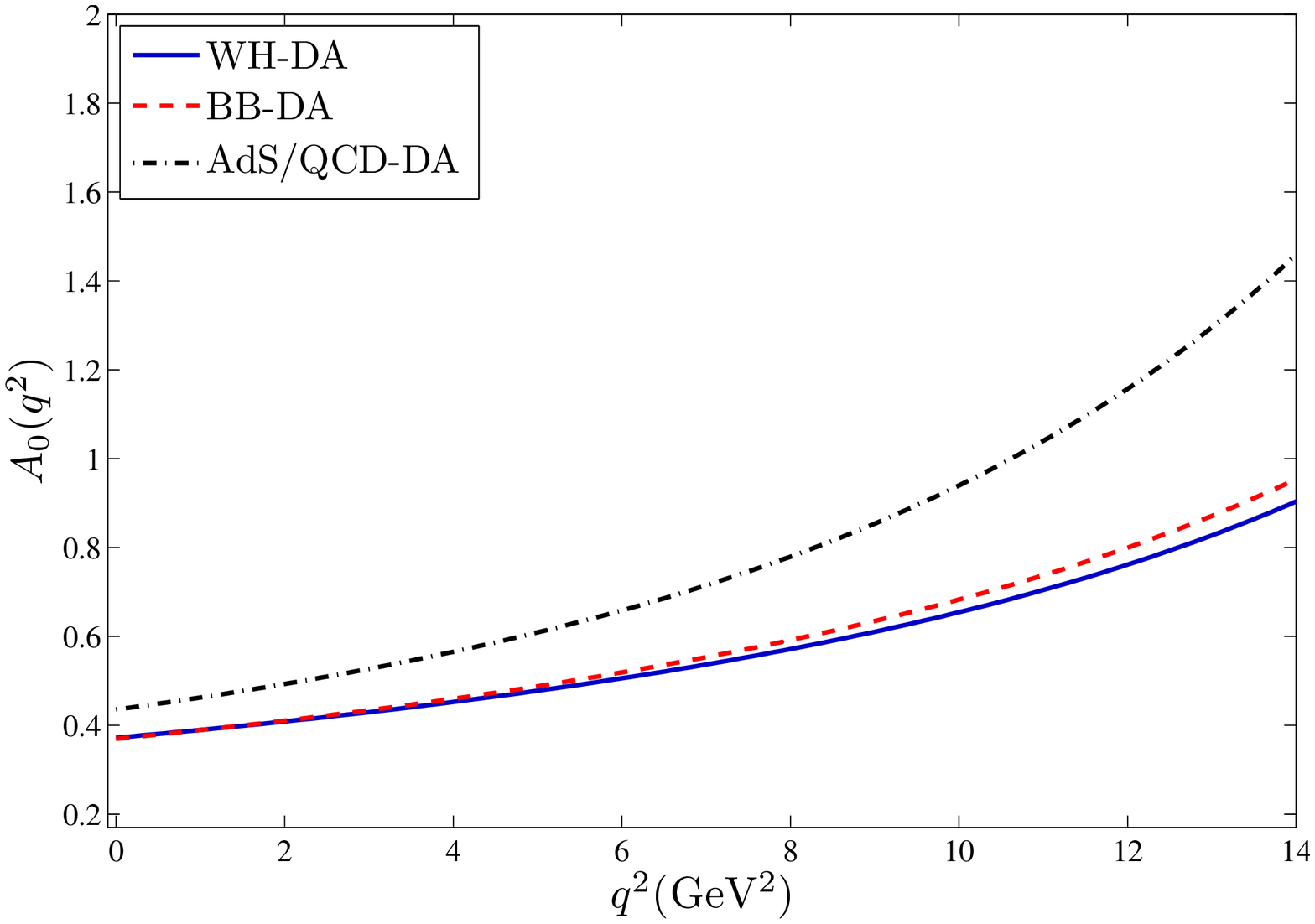}
\includegraphics[width=0.24\textwidth]{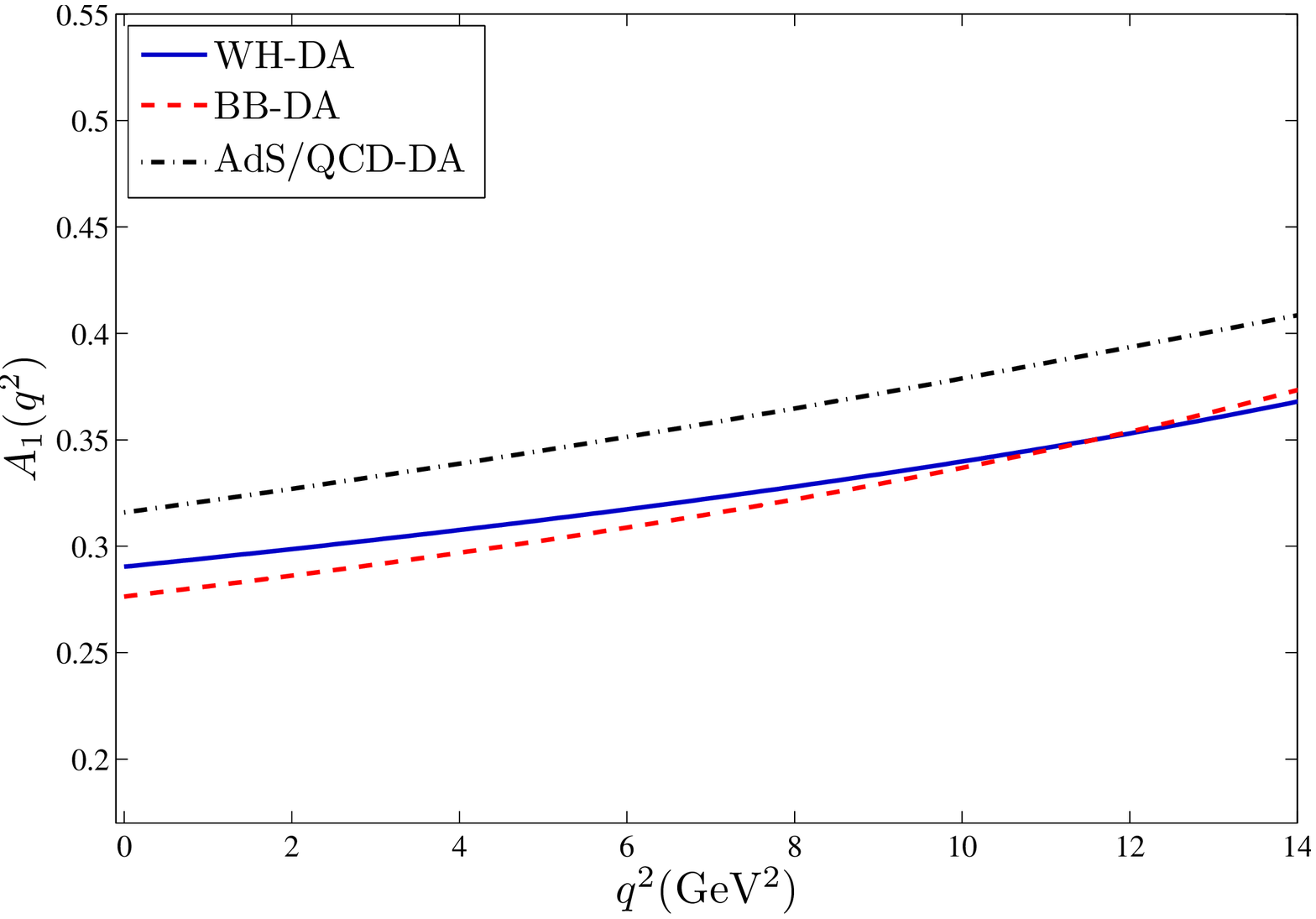}
\includegraphics[width=0.24\textwidth]{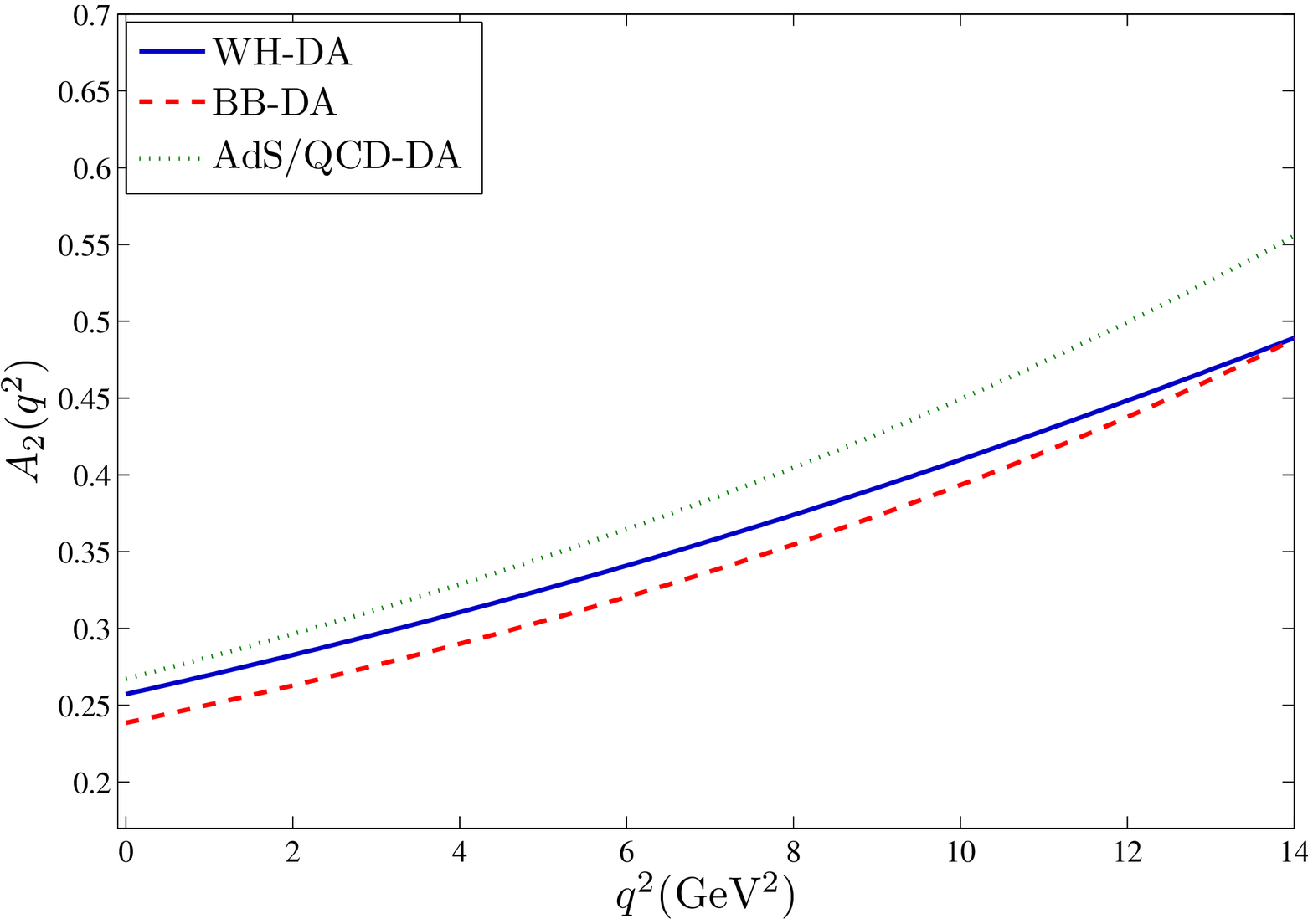}
\includegraphics[width=0.24\textwidth]{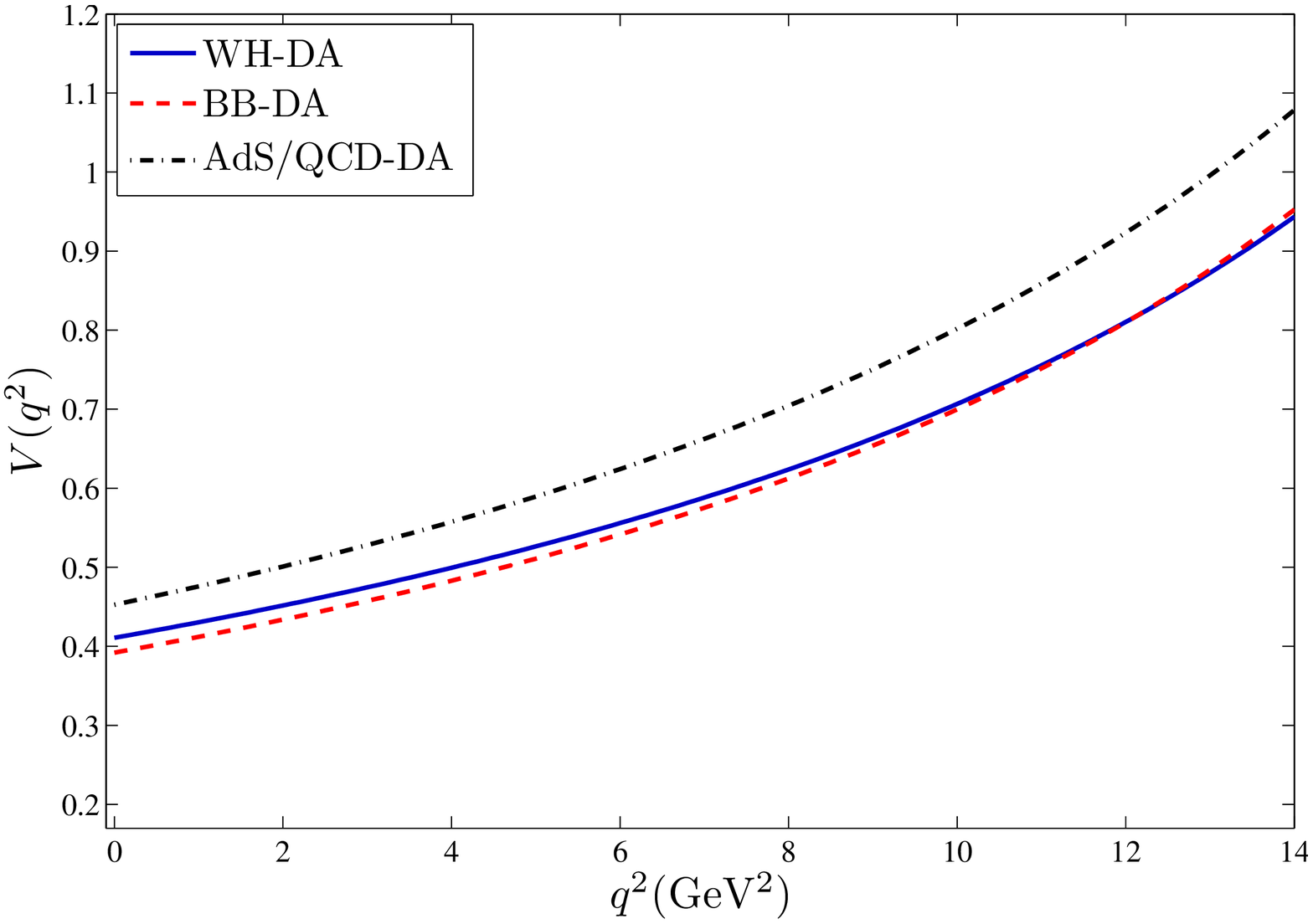}
\caption{A comparison of LCSRs of the TFFs $A_{0,1,2}(q^2)$ and $V(q^2)$ under different choice of $\phi_{2;{K^*}}^\bot$.} \label{TFF:A1A2Vcom}
\end{center}
\end{figure*}

\begin{figure*}
\begin{center}
\includegraphics[width=0.24\textwidth]{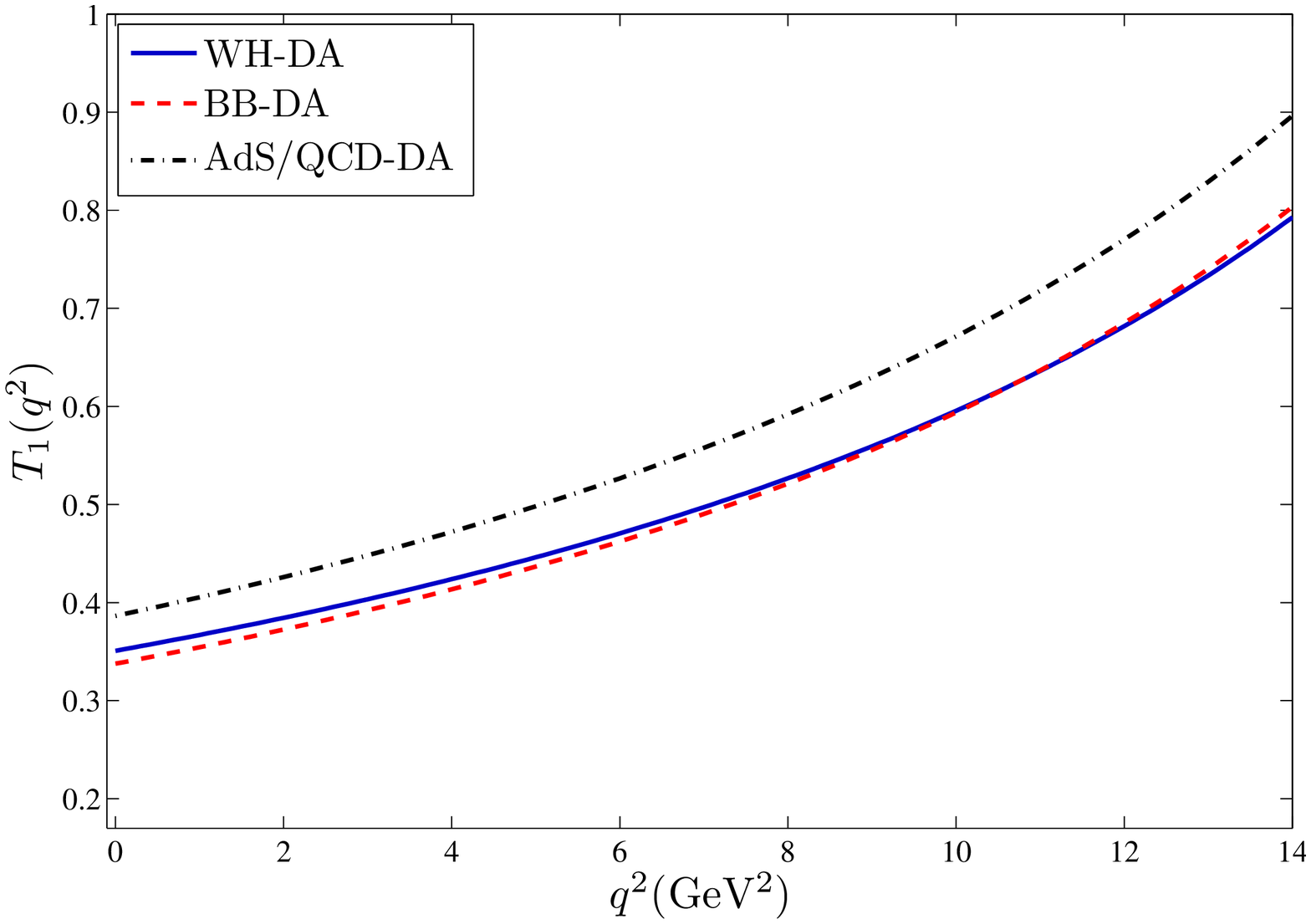}
\includegraphics[width=0.24\textwidth]{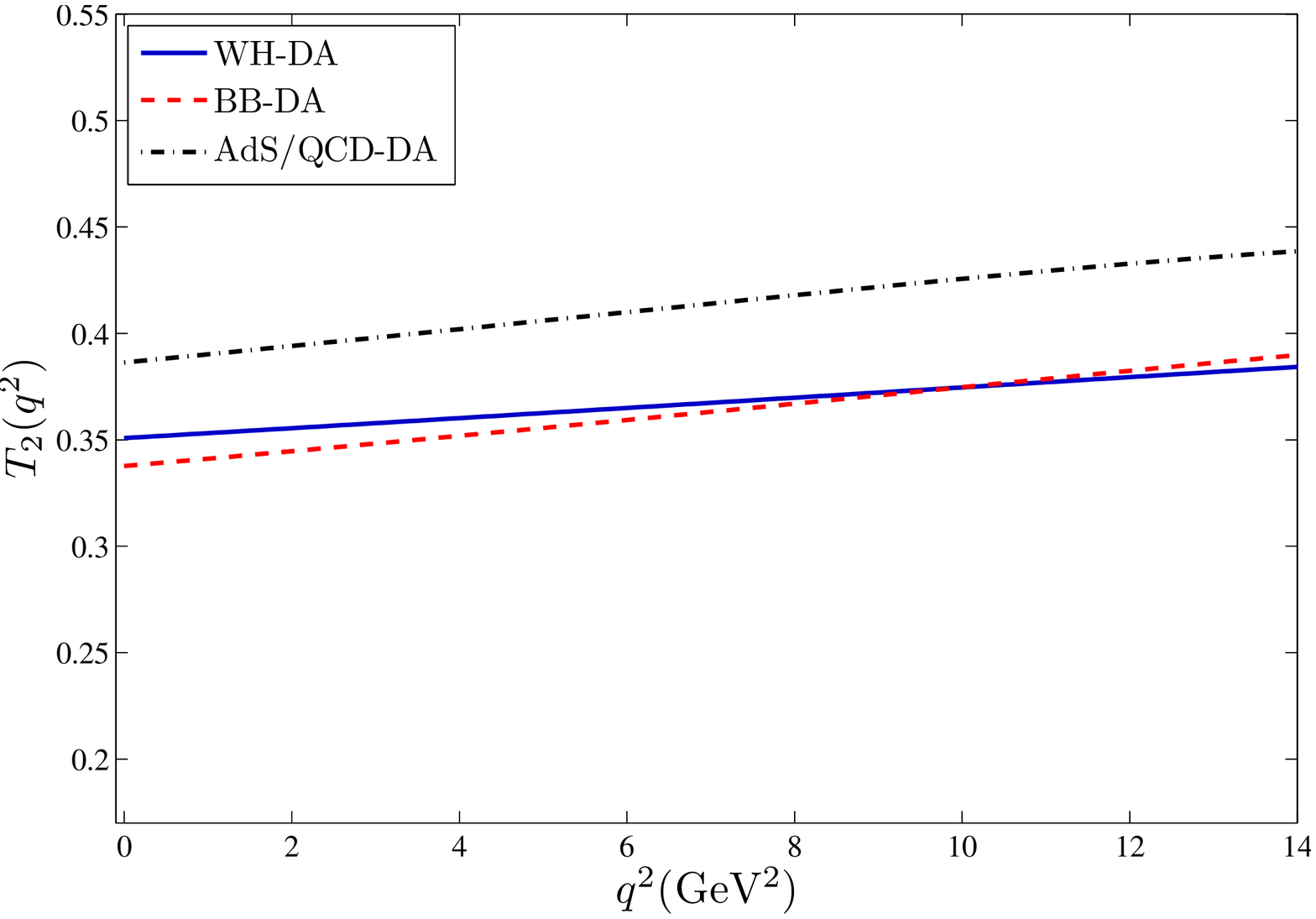}
\includegraphics[width=0.24\textwidth]{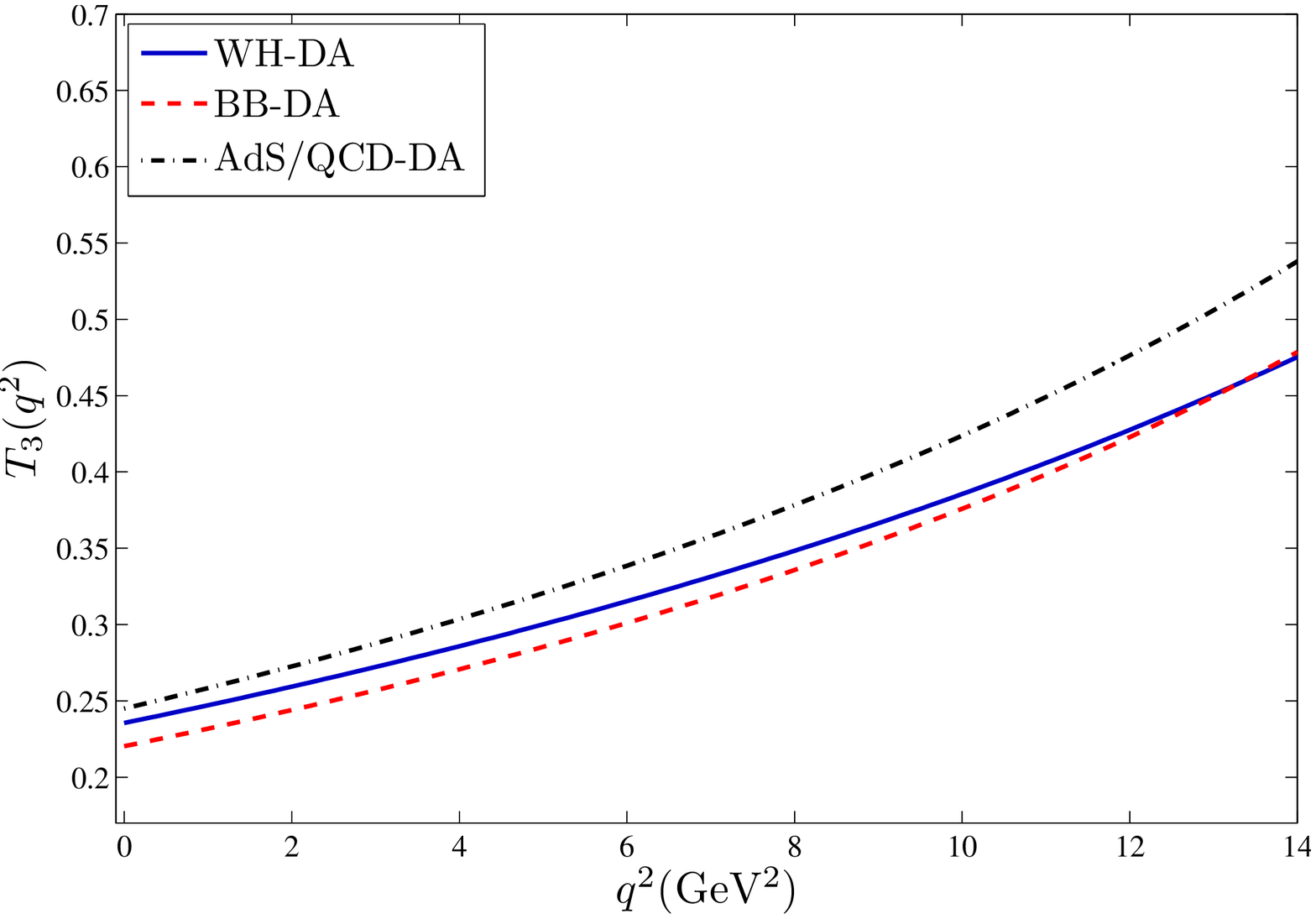}
\includegraphics[width=0.24\textwidth]{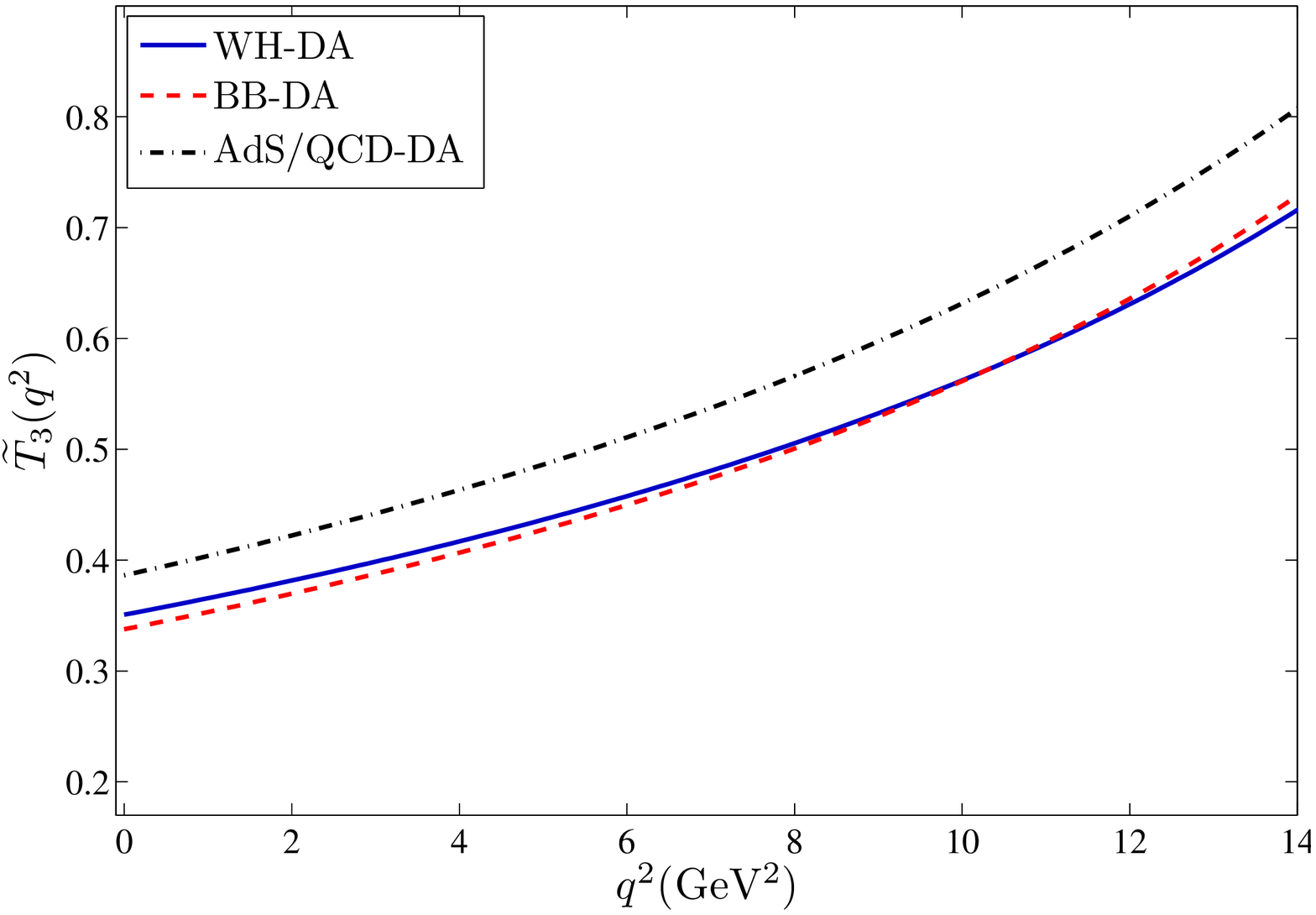}
\caption{A comparison of LCSRs of the TFFs $T_{1,2,3}(q^2)$ and $\widetilde T_3$ under different choice of $\phi_{2;{K^*}}^\bot$.} \label{TFF:T1T2T3com}
\end{center}
\end{figure*}

As a final remark, to have a clear look at how the twist-2 LCDA $\phi_{2;K^*}^\bot$ affects the TFFs, we calculate the LCSRs (\ref{TFF_A1}--\ref{TFF_T3}) under three different models for $\phi_{2;K^*}^\bot$, i.e. the WH-DA, the BB-DA and the AdS/QCD-DA. All other inputs are taken to be the same. The results are presented in Figs.(\ref{TFF:A1A2Vcom}, \ref{TFF:T1T2T3com}). The shapes of the TFFs for the WH-DA and BB-DA are close to each other, while the ones of the AdS/QCD-DA are quite different. This is reasonable, since the WH-DA and the BB-DA have the same Gegenbauer moments.

\subsection{An extrapolation of the TFFs to high $q^2$-region}

\begin{table}[htb]
\begin{tabular}{ c  c c c c c c c }
\hline
&$A_1$ & $A_2$ & $V$   & $A_0$  & $T_1$  & $T_2$  & $T_3$
\\ \hline
$a_1^i$   & 0.988 & $-0.456$ & $-0.644$ & $-0.523$ & $-0.461$  & 1.578 & $-0.244$
\\
$a_2^i$   & 0.432 & $-5.048$ & $-1.140$ &  1.370   & $-0.692$  & 1.590 & $-3.130$
\\
$\Delta$  &  0.07 &   0.35   & 0.04     &  0.48    & 0.04      & 0.22  & 0.10
\\ \hline
\end{tabular}
\caption{The fitted parameters $a^i_{1,2}$ for the $B\to K^*$ TFFs $F_i$, in which all the LCSR parameters are set to be their central values. $\Delta$ is the quality of fit.} \label{analytic}
\end{table}

In principal, the LCSRs for the $B\to K^*$ TFFs are applicable in small and intermediate $q^2$-region, e.g. $0\leq q^2 \leq 14{\rm GeV}^2$. In order to facilitate the applicability of the obtained LCSRs, as suggested by Refs.~\cite{Khodjamirian:2010vf, Bourrely:2008za}, we perform fits of the full analytical result to a simplified series expansion, which is based on a rapidly converging series over the parameter
\begin{eqnarray}
z(t)=\frac{\sqrt{t_+ - t}-\sqrt{t_+ - t_0}}{\sqrt{t_+ - t}+\sqrt{t_+ - t_0}},
\end{eqnarray}
where $t_\pm=(m_B\pm m_V)^2$ and $t_0=t_+(1-\sqrt{1-t_-/t_+})$. Then we expand the form factors as
\begin{eqnarray}
F_i(q^2)=\frac1{1-q^2/m_{R,i}^2}\sum_{k=0,1,2} a_k^i [z(q^2)-z(0)]^k , \label{extrapolation}
\end{eqnarray}
where $F_i$ stands for the TFFs $A_{0,1,2}$, $V$ or $T_{1,2,3}$, respectively. The values of the resonance masses $m_{R,i}$ can be found in Ref.~\cite{Straub:2015ica}. The coefficients $a_0^i$ is defined as $F_i(0)$. Then the first term of the expansion (\ref{extrapolation}) is the usually adopted single-pole extrapolation formulae, $F_i(q^2)=F_i(0)/(1-q^2/m_{R,i}^2)$. The parameters $a_1^i$ and $a_2^i$ are determined by requiring the ``quality'' of fit ($\Delta$) to be less than one~\cite{Fu:2014pba, Ball:2004rg}. Here $\Delta$ is defined as
\begin{equation}
\Delta=\frac{\sum_t\left|F_i(t)-F_i^{\rm fit}(t)\right|} {\sum_t\left|F_i(t)\right|}\times 100, \label{delta}
\end{equation}
where $t\in[0,\frac{1}{2},\cdots,\frac{27}{2},14]{\rm GeV}^2$. As an illustration, we put the values of the fitting parameters $a^i_{1,2}$ in Table~\ref{analytic}, in which all the LCSR parameters are set to be their central values.

\begin{figure*}
\begin{center}
\includegraphics[width=0.24\textwidth]{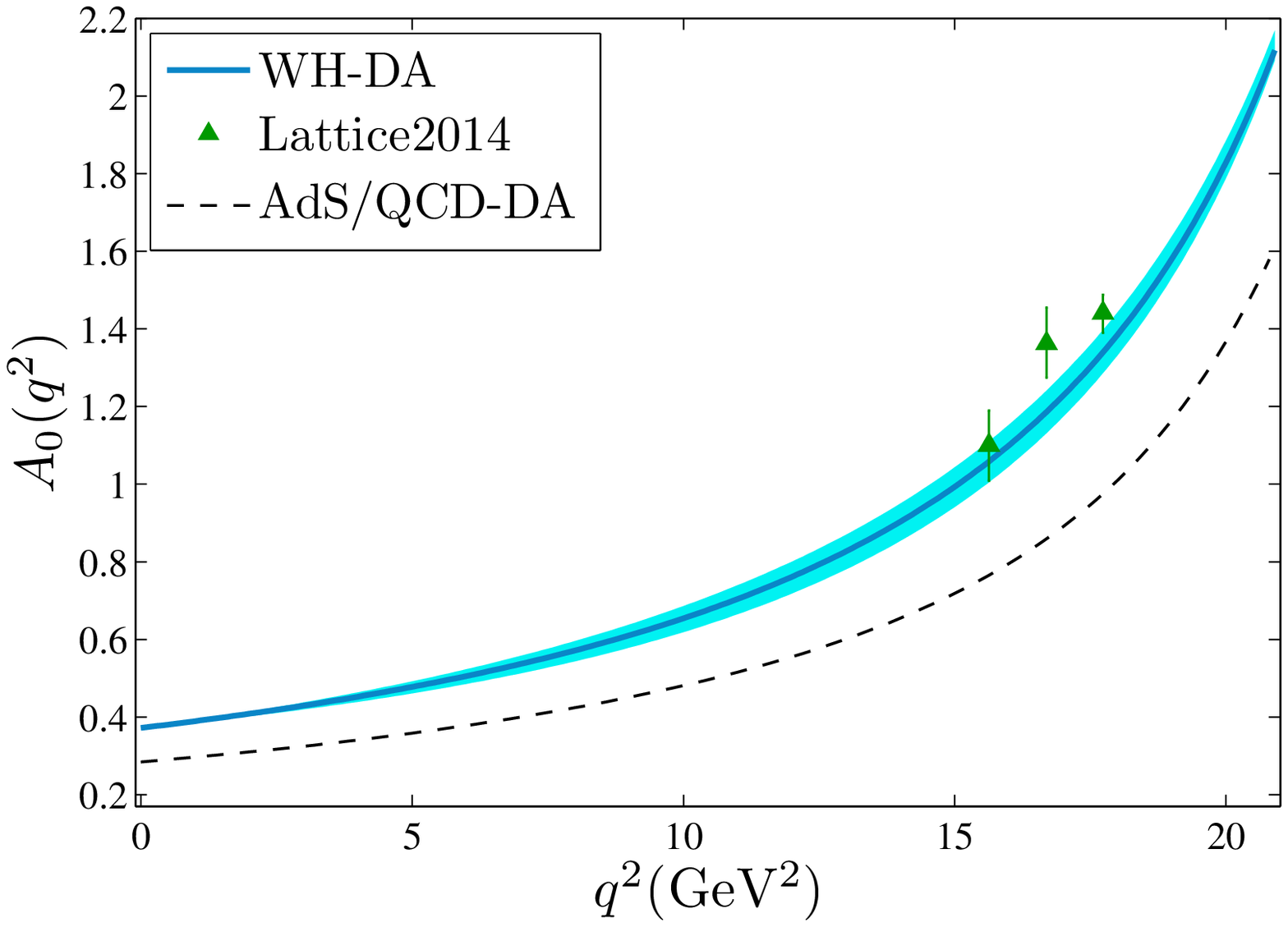}
\includegraphics[width=0.24\textwidth]{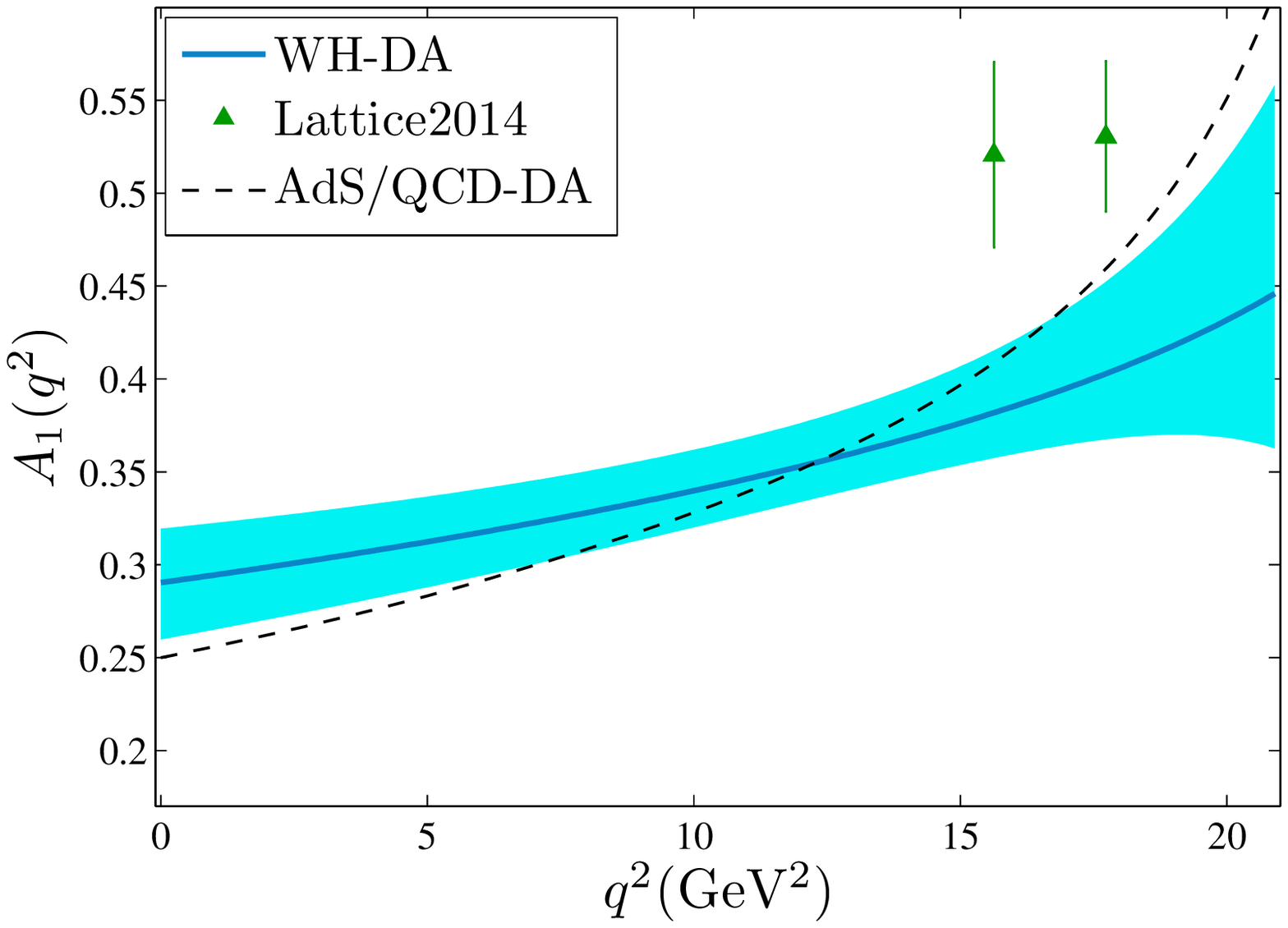}
\includegraphics[width=0.24\textwidth]{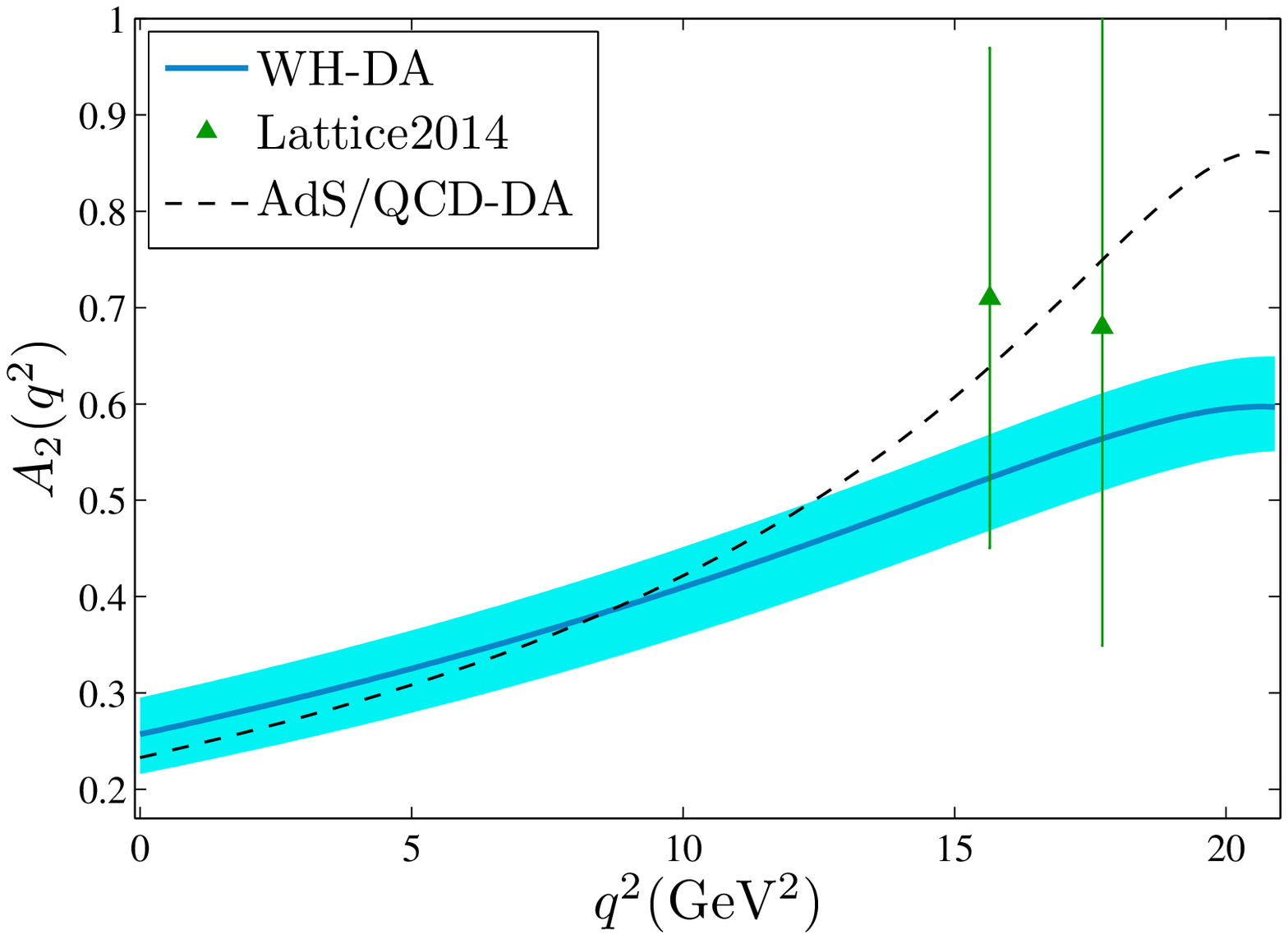}
\includegraphics[width=0.24\textwidth]{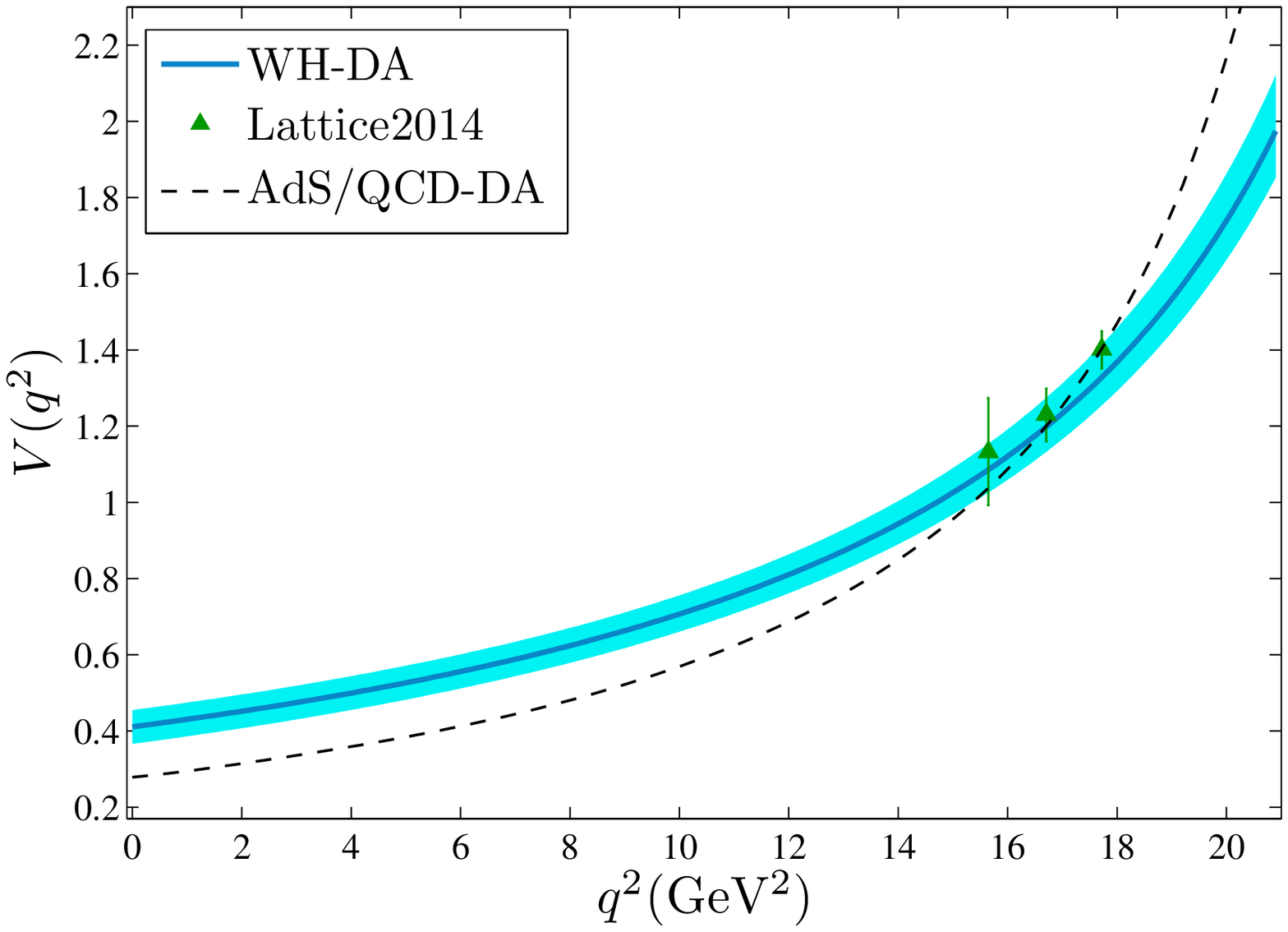}
\caption{The extrapolated $B\to K^*$ axial-vector and vector TFFs $A_{0,1,2}(q^2)$ and $V(q^2)$. As a comparison, the lattice QCD~\cite{Horgan:2013hoa} and AdS/QCD~\cite{Ahmady:2014sva} predictions are also presented. } \label{TFF:A1A2V}
\end{center}
\end{figure*}

\begin{figure*}
\begin{center}
\includegraphics[width=0.24\textwidth]{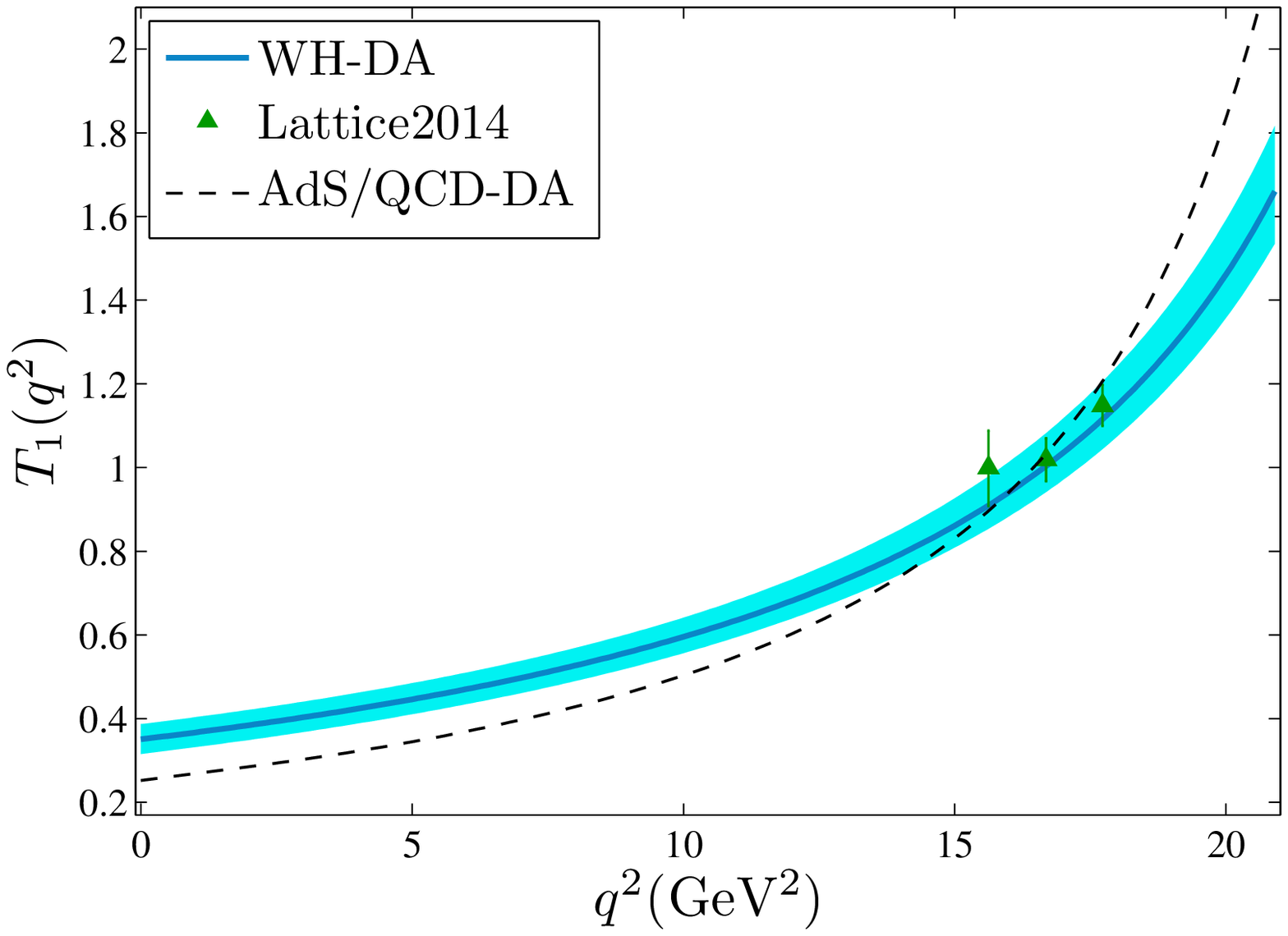}
\includegraphics[width=0.24\textwidth]{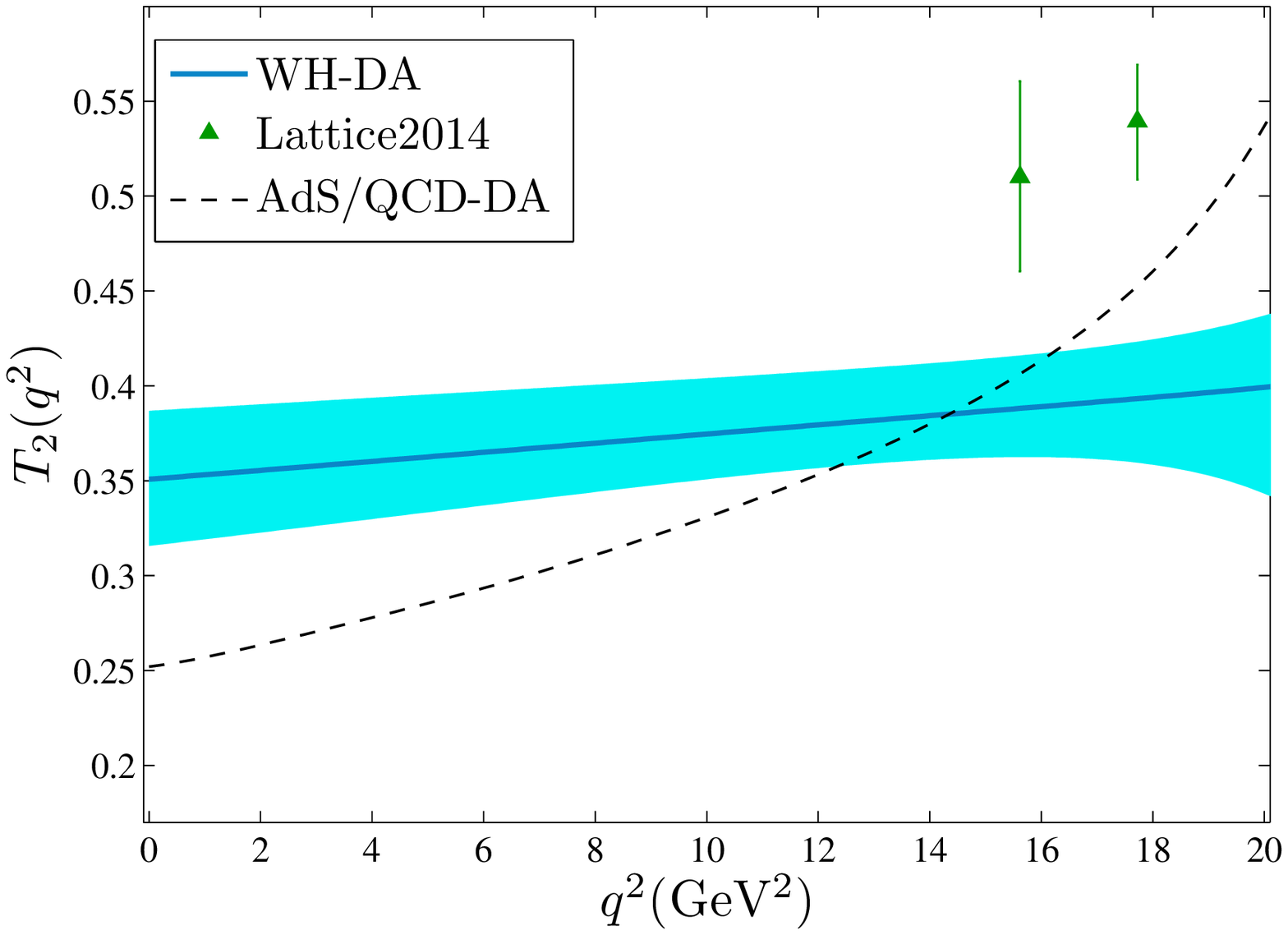}
\includegraphics[width=0.24\textwidth]{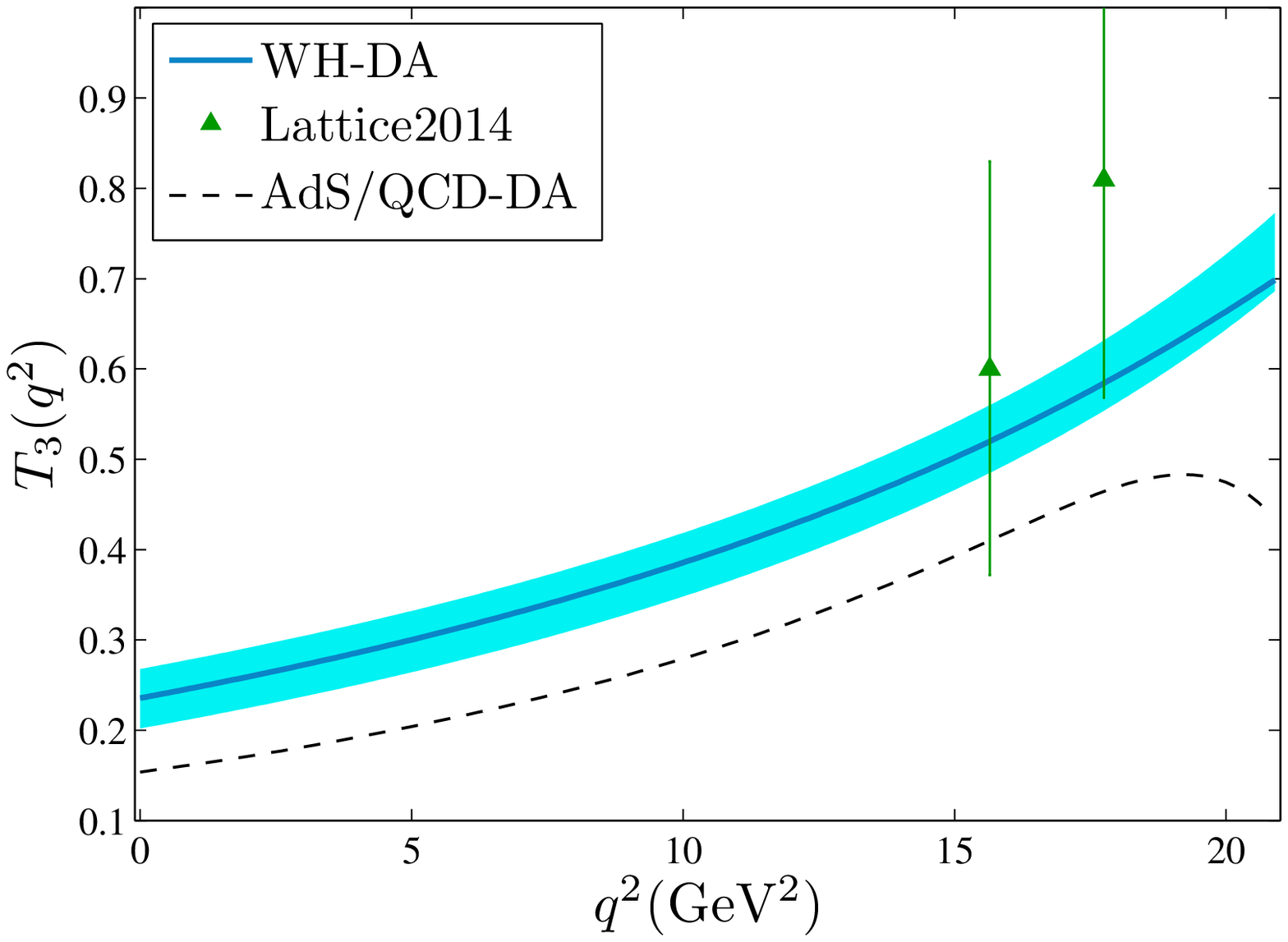}
\includegraphics[width=0.24\textwidth]{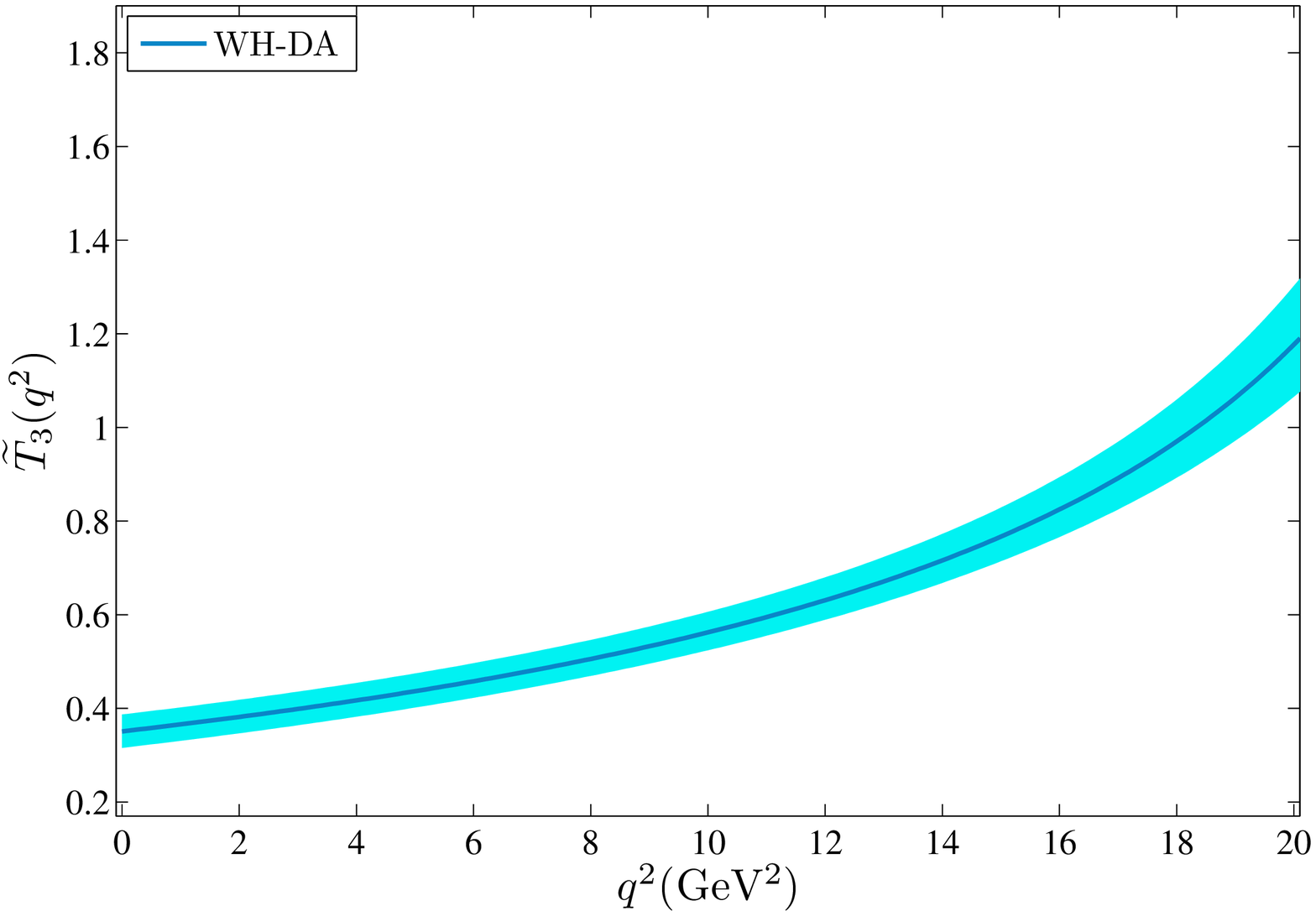}
\caption{The extrapolated $B\to K^*$ tensor TFFs $T_{1,2,3}(q^2)$ and $\widetilde T_3$. As a comparison, the lattice QCD~\cite{Horgan:2013hoa} and AdS/QCD~\cite{Ahmady:2014sva} predictions are also presented. } \label{TFF:T1T2T3}
\end{center}
\end{figure*}

We put the extrapolated $B\to {K^*}$ TFFs $A_{0,1,2}(q^2)$, $V(q^2)$ and $T_{1,2,3}(q^2)$ in Figs.(\ref{TFF:A1A2V}, \ref{TFF:T1T2T3}), in which the available lattice QCD predictions~\cite{Horgan:2013hoa} and the AdS/QCD predictions~\cite{Ahmady:2014sva} are also included as a comparison. The shaded band in Figs.(\ref{TFF:A1A2V}, \ref{TFF:T1T2T3}) stands for the squared average of the uncertainties from $a^{i}_{1,2}$, and the choices of $\mu$, $s_0$ and $M^2$. Figs.(\ref{TFF:A1A2V}, \ref{TFF:T1T2T3}) show that all the TFFs are in good agreement with the lattice predictions within errors, except for the two TFFs $A_1$ and $T_2$.

\subsection{The branching fraction of $B \to K^* \mu^+ \mu^-$}

As an application, we apply the extrapolated $B\to K^*$ TFFs to study the branching fraction of the dileptonic decay $B \to K^* \mu^+ \mu^-$. This decay is very useful for precise tests of the standard model and for probing new physics beyond standard model. Its differential branching fraction can be written as~\cite{Aliev:1996hb}
\begin{widetext}
\begin{eqnarray}
\frac{d{\cal B}}{dq^2} &=& \tau_B \frac{G_F^2 \alpha^2}{2^{13}\pi^5} \frac{|V_{tb}V_{ts}^*|^2 \sqrt\lambda v}{3m_B} \bigg\{ (2m_\mu^2 + m_B^2 s)\Big[ 16(|A{|^2} + |C|^2) m_B^4\lambda  + 2(|B_1|^2 + |D_1|^2) \nonumber\\
&& \times \frac{\lambda  + 12rs}{rs} + 2(|B_2|^2 + |D_2|^2)\frac{m_B^4\lambda^2}{rs} - 4[\Re {\rm{e}}(B_1 B_2^*) + \Re {\rm{e}}(D_1 D_2^*)]\frac{m_B^2\lambda }{rs}(1-r-s)\Big]\nonumber\\
&&  + 6m_\mu ^2\Big[ - 16|C|^2m_B^4\lambda  + 4\Re {\rm{e}}(D_1 D_3^*)\frac{m_B^2\lambda}{r} - 4\Re {\rm{e}}(D_2D_3^*)\frac{m_B^4(1 - r)\lambda }{r} + 2|D_3|^2\frac{m_B^4s\lambda }{r}\nonumber\\
&&   - 4\Re {\rm{e}}(D_1D_2^*)\frac{m_B^2\lambda }{r} - 24|D_1|^2 + 2|D_2|^2\frac{m_B^4\lambda}{r}(2 + 2r - s)\Big] \bigg\}\label{eq:dB}
\end{eqnarray}
\end{widetext}
where $r=m_{K^*}^2/m_B^2$, $s=q^2/m_B^2$, and the phase-space factor $\lambda=1+r^2+s^2-2r-2s-2rs$. The muon velocity $v=(1-4 m_{\mu}^2/q^2)^{1/2}$, where $m_{\mu}$ is muon mass. $\tau_B$ is the average lifetime of those of $B^0$- and $B^+$-mesons~\cite{Agashe:2014kda}. The coefficients $A$, $B_{1,2}$, $C$ and $D_{1,2,3}$ are functions of $B\to K^*$ TFFs and the Wilson coefficients, whose explicit forms can be read from Ref.\cite{Aliev:1996hb}. At the low and high $q^2$-region, we may need more effects to achieve a reliable prediction~\cite{Beneke:2001at, Grinstein:2004vb}. Especially the next-to-leading order QCD corrections to the matrix elements of current-current operators improve the values of the Wilson coefficients and can enhance the rate at high $q^2$ to a certain degree~\cite{Grinstein:2004vb}. Here, to concentrate our attention on the effects of the $B\to K^*$ TFFs, we shall directly adopt Eq.(\ref{eq:dB}) to do our analysis without considering those subtle points.

\begin{figure}[htb]
\begin{center}
\includegraphics[width=0.4\textwidth]{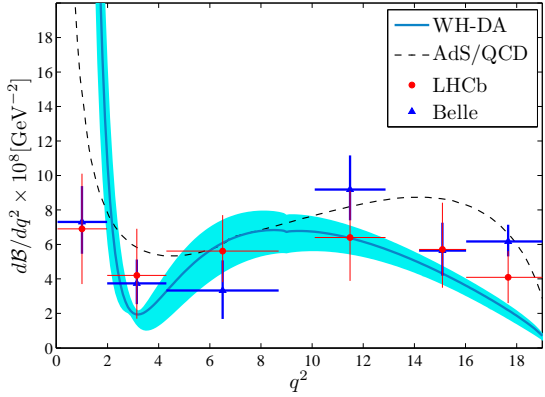}
\caption{The differential branching fraction $d{\cal B}/dq^2(B\to K^*\mu^+\mu^-) $ as a function of $q^2$ under WH-DA model, which the theoretical uncertainties are included. The Belle data~\cite{Wei:2009zv}, the LHCb data~\cite{Aaij:2012cq, Aaij:2013qta, LHCb:2012aja}, and the AdS/QCD prediction~\cite{Ahmady:2014sva} are included as a comparison.} \label{dB}
\end{center}
\end{figure}

\begin{figure}[htb]
\begin{center}
\includegraphics[width=0.4\textwidth]{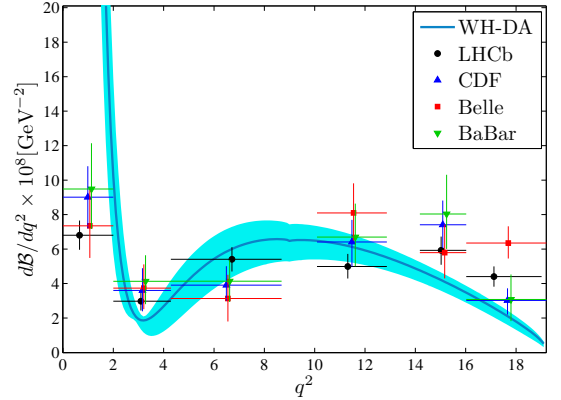}
\caption{The differential branching fraction $d{\cal B}/dq^2(B^0\to K^{*0}\mu^+\mu^-)$ as a function of $q^2$. The LHCb data~\cite{LHCb:2012aja}, the BABAR data~\cite{Akar:2014hda}, the Belle data~\cite{Wei:2009zv}, and the CDF data~\cite{Aaltonen:2011ja} are included as a comparison. } \label{dB_B0}
\end{center}
\end{figure}

Our predictions for the differential branching fraction of $B \to K^* \mu^+ \mu^-$ are presented in Fig.~\ref{dB}, where the LHCb data~\cite{Aaij:2012cq, Aaij:2013qta, LHCb:2012aja} and the AdS/QCD prediction~\cite{Ahmady:2014sva} are included as a comparison. For the LHCb data, we have adopted the weighted average of the branching fractions for $B^+\to K^{*+}\mu^+\mu^-$~\cite{Aaij:2012cq,Aaij:2013qta} and $B^0\to K^{*0}\mu^+\mu^-$~\cite{LHCb:2012aja}.

Experimentally, the charged $K^{*+}$ and the neutral $K^{*0}$ can be separated, then one can distinguish those two decay channels. As a useful reference, we also present the differential branching fraction for $B^0\to K^{*0}\mu^+\mu^-$ in Fig.~\ref{dB_B0}, which has been measured by the LHCb, the BABAR, the Belle and the CDF collaborations~\cite{LHCb:2012aja, Akar:2014hda, Wei:2009zv, Aaltonen:2011ja}. In those two figures, the shaded bands are theoretical uncertainties. They show our present predictions are consistent with the LHCb data within errors, especially for larger $q^2$-region, $q^2 > 2 {\rm GeV}^2$.

By integrating the differential branching fraction Eq.(\ref{eq:dB}) over all allowable $q^2$-region, we obtain
\begin{eqnarray}
{\cal B} =\left(1.088^{+0.261}_{-0.205}\right)\times 10^{-6},
\end{eqnarray}
where the errors are squared averages of the errors caused by varying $s_0$, $M^2$, $m_b$, and $\mu$ within the determined/adopted ranges shown in Sec.III.A and by taking the $B$-meson lifetimes $\tau(B^0)=1.519\pm0.005$ ps and $\tau(B^\pm)=1.638\pm0.004$ps~\cite{Agashe:2014kda} \footnote{Here, we adopt the central values for all the Wilson coefficients within the SM~\cite{Altmannshofer:2008dz} to do the calculation, and the uncertainties of them have not been taken into consideration.}. It is found that the predicted total branching fraction shows a better agreement with the LHCb measurements~\cite{Aaij:2012cq, Aaij:2013qta, LHCb:2012aja}, which also agrees with a pQCD prediction by including the ${\cal O}(\alpha_s)$ and $\Lambda_{\rm QCD}/m_b$ corrections, i.e. ${\cal B}(B\to K^*\mu^+\mu^-)=(1.19\pm0.39)\times 10^{-6}$~\cite{Ali:2002jg}. Apart from the TFFs, there are many other sources of theoretical uncertainties for this process which are still in progress, cf.Refs.\cite{Descotes-Genon:2015hea, Niehoff:2015qda, Ahmady:2015fha, Sahoo:2015qha, Descotes-Genon:2015xqa, Hofer:2015kka, Descotes-Genon:2014joa}. When the precision of those sources have been improved in the future, we may finally get a definite conclusion on the $K^*$-meson transverse leading-twist LCDA $\phi_{2;K^*}^\bot$.

\section{Summary} \label{summary}

We have recalculated the $B\to K^*$ TFFs by using the chiral LCSR approach and by taking the $K^*$ meson ${\rm SU}_{f}(3)$-breaking effect into consideration.

By taking the chiral correlator to do the LCSR calculation, it has been found that the contributions from the mostly uncertain high-twist LCDAs can be greatly suppressed. As required, the LCSRs (\ref{TFF_A1}--\ref{TFF_T3}) show that the chiral-even LCDAs $\phi_{2;{K^*}}^\|$, $\phi_{3;{K^*}}^\bot$, $\psi_{3;{K^*}}^\bot$, $\Phi_{3;{K^*}}^\|$ and $\widetilde\Phi_{3;{K^*} }^\bot$ provide zero contributions to the LCSRs; all the remaining non-zero twist-3 and twist-4 LCDAs are $\delta^2$-suppressed and can only provide less than $10\%$ contributions to the LCSRs. Thus more accurate LCSRs for the $B\to K^*$ TFFs have been achieved, which inversely provide good platforms for testing the properties of the transverse leading-twist LCDA $\phi_{2;K^*}^\bot$.

To compare with the lattice QCD predictions on the $B\to K^*$ TFFs, we have suggested a convenient model (\ref{DA_WH}) for $\phi_{2;K^*}^\bot$, in which two parameters $B_{2;K^*}^\bot$ and $C_{2;K^*}^\bot$ dominantly control its longitudinal behavior. By comparing with the lattice QCD predictions, we observe that apart from $A_1$ and $T_2$, other TFFs show good agreement with the lattice QCD prediction, especially for the $A_0$, $V$, $T_1$, and $T_3$. The twist-2 LCDA $\phi_{2;K^*}^\bot$ has a small asymmetry due to the $K^*$-meson ${\rm SU}_f(3)$-breaking effect, which is controlled by the parameter $B_{2;K^*}^\bot$. In this paper, the central value of $B_{2;K^*}^\bot = 0.0038$ indicates a small asymmetry. Meanwhile, the parameter $C_{2;K^*}^\bot$ controls the shape of WH-DA, determining wether it is double-peak or single-peak. A smaller $C_{2;K^*}^\bot$ indicates a single-peak behavior and a larger one indicates a double-peak behavior. At present, the lattice QCD predictions are of large errors, and a more accurate lattice prediction shall lead to a better constraint on $\phi_{2;{K^*}}^\bot$.

As an application of the obtained $B\to K^*$ TFFs, we have further predicted the differential branching fraction of the decay $B \to K^* \mu^+ \mu^-$. The predicted branching fractions are consistent with the LHCb and the Belle measurements within errors, especially for the intermediate and large $q^2$-region, $q^2 > 2 {\rm GeV}^2$. After integrating the differential branching fraction (\ref{eq:dB}) over all allowable $q^2$-region, the integrated branching fraction ${\cal B}(B\to K^*\mu^+\mu^-)$ also shows a better agreement with the LHCb measurements~\cite{Aaij:2012cq, Aaij:2013qta, LHCb:2012aja} within errors. \\

{\bf Acknowledgments}:  This work was supported in part by Natural Science Foundation of China under Grant No.11275280, by the Fundamental Research Funds for the Central Universities under Grant No.CDJZR305513.


\begin{thebibliography}{99}

\bibitem{Descotes-Genon:2013wba}
  S.~Descotes-Genon, J.~Matias and J.~Virto,
  ``Understanding the $B\to K^* \mu^+ \mu^-$ Anomaly,''
  Phys.\ Rev.\ D {\bf 88}, 074002 (2013).

\bibitem{Bobeth:2010wg}
  C.~Bobeth, G.~Hiller and D.~van Dyk,
  ``The Benefits of $\bar{B} \to \bar{K}^* l^+ l^-$ Decays at Low Recoil,''
  JHEP {\bf 1007}, 098 (2010).

\bibitem{Alok:2010zd}
  A.~K.~Alok, A.~Datta, A.~Dighe, M.~Duraisamy, D.~Ghosh and D.~London,
  ``New Physics in $b \to s \mu^+ \mu^-$: CP-Conserving Observables,''
  JHEP {\bf 1111}, 121 (2011).

\bibitem{Alok:2011gv}
  A.~K.~Alok, A.~Datta, A.~Dighe, M.~Duraisamy, D.~Ghosh and D.~London,
  ``New Physics in $b \to s \mu^+ \mu^-$: CP-Violating Observables,''
  JHEP {\bf 1111}, 122 (2011).

\bibitem{DescotesGenon:2011yn}
  S.~Descotes-Genon, D.~Ghosh, J.~Matias and M.~Ramon,
  ``Exploring New Physics in the $C_7-C_7'$ plane,''
  JHEP {\bf 1106}, 099 (2011).

\bibitem{Bobeth:2011gi}
  C.~Bobeth, G.~Hiller and D.~van Dyk,
  ``More Benefits of Semileptonic Rare $B$ Decays at Low Recoil: CP Violation,''
  JHEP {\bf 1107}, 067 (2011).

\bibitem{Becirevic:2011bp}
  D.~Becirevic and E.~Schneider,
  ``On transverse asymmetries in $B \to K^* l^+ l^-$,''
  Nucl.\ Phys.\ B {\bf 854}, 321 (2012).

\bibitem{Altmannshofer:2011gn}
  W.~Altmannshofer, P.~Paradisi and D.~M.~Straub,
  ``Model-Independent Constraints on New Physics in $b \to s$ Transitions,''
  JHEP {\bf 1204}, 008 (2012).

\bibitem{Bobeth:2011nj}
  C.~Bobeth, G.~Hiller, D.~van Dyk and C.~Wacker,
  ``The Decay $B \to K \ell^+ \ell^-$ at Low Hadronic Recoil and Model-Independent $\Delta B = 1$ Constraints,''
  JHEP {\bf 1201}, 107 (2012).

\bibitem{Matias:2012xw}
  J.~Matias, F.~Mescia, M.~Ramon and J.~Virto,
  ``Complete Anatomy of $\bar{B}_d \to \bar{K}^{* 0} (\to K \pi)l^+l^-$ and its angular distribution,''
  JHEP {\bf 1204}, 104 (2012).

\bibitem{Beaujean:2012uj}
  F.~Beaujean, C.~Bobeth, D.~van Dyk and C.~Wacker,
  ``Bayesian Fit of Exclusive $b \to s \bar\ell\ell$ Decays: The Standard Model Operator Basis,''
  JHEP {\bf 1208}, 030 (2012).

\bibitem{Altmannshofer:2008dz}
  W.~Altmannshofer, P.~Ball, A.~Bharucha, A.~J.~Buras, D.~M.~Straub and M.~Wick,
  ``Symmetries and Asymmetries of $B \to K^{*} \mu^{+} \mu^{-}$ Decays in the Standard Model and Beyond,''
  JHEP {\bf 0901}, 019 (2009).


\bibitem{Altmannshofer:2012az}
  W.~Altmannshofer and D.~M.~Straub,
  ``Cornering New Physics in $b \to s$ Transitions,''
  JHEP {\bf 1208}, 121 (2012).

\bibitem{Becirevic:2012dx}
  D.~Becirevic, E.~Kou, A.~Le Yaouanc and A.~Tayduganov,
  ``Future prospects for the determination of the Wilson coefficient $C_{7\gamma}^\prime$,''
  JHEP {\bf 1208}, 090 (2012).

\bibitem{DescotesGenon:2012zf}
  S.~Descotes-Genon, J.~Matias, M.~Ramon and J.~Virto,
  ``Implications from clean observables for the binned analysis of $B \to K*\mu^+\mu^-$ at large recoil,''
  JHEP {\bf 1301}, 048 (2013).

\bibitem{Bobeth:2012vn}
  C.~Bobeth, G.~Hiller and D.~van Dyk,
  ``General analysis of $\bar{B} \to \bar{K}^{(*)}\ell^+ \ell^-$  decays at low recoil,''
  Phys.\ Rev.\ D {\bf 87}, 034016 (2013).

\bibitem{Descotes-Genon:2013vna}
  S.~Descotes-Genon, T.~Hurth, J.~Matias and J.~Virto,
  ``Optimizing the basis of $B\to K^* l l$ observables in the full kinematic range,''
  JHEP {\bf 1305}, 137 (2013).


\bibitem{Aaltonen:2011ja}  T.~Aaltonen {\it et al.} [CDF Collaboration],
  ``Measurements of the Angular Distributions in the Decays $B \to K^{(*)} \mu^+ \mu^-$ at CDF,''
  Phys.\ Rev.\ Lett.\  {\bf 108}, 081807 (2012).

\bibitem{Aaltonen:2011qs}
  T.~Aaltonen {\it et al.} [CDF Collaboration],
  ``Observation of the Baryonic Flavor-Changing Neutral Current Decay $\Lambda_{b} \to \Lambda \mu^{+} \mu^{-}$,''
  Phys.\ Rev.\ Lett.\  {\bf 107}, 201802 (2011).

\bibitem{Lees:2012tva}
  J.~P.~Lees {\it et al.} [BaBar Collaboration],
  ``Measurement of Branching Fractions and Rate Asymmetries in the Rare Decays $B \to K^{(*)} l^+ l^-$,''
  Phys.\ Rev.\ D {\bf 86}, 032012 (2012).

\bibitem{Wei:2009zv}
  J.~T.~Wei {\it et al.} [Belle Collaboration],
  ``Measurement of the Differential Branching Fraction and Forward-Backword Asymmetry for $B\to K^{(*)} l^+ l^-$,''
  Phys.\ Rev.\ Lett.\  {\bf 103}, 171801 (2009).

\bibitem{Aaij:2012cq}
  R.~Aaij {\it et al.} [LHCb Collaboration],
  ``Measurement of the isospin asymmetry in $B \to K^{(*)}\mu^+\mu^-$ decays,''
  JHEP {\bf 1207}, 133 (2012).

\bibitem{Aaij:2013qta}
  R.~Aaij {\it et al.} [LHCb Collaboration],
  ``Measurement of Form-Factor-Independent Observables in the Decay $B^{0} \to K^{*0} \mu^+ \mu^-$,''
  Phys.\ Rev.\ Lett.\  {\bf 111}, 191801 (2013).


\bibitem{LHCb:2012aja}
  T. Blake and N. Serra [LHCb Collaboration],
  ``Differential branching fraction and angular analysis of the $B^{0} \to K^{*0} \mu^{+}\mu^{-}$ decay,''
  LHCb-CONF-2012-008, CERN-LHCb-CONF-2012-008.

\bibitem{Aaij:2013iag}
  R.~Aaij {\it et al.} [LHCb Collaboration],
  ``Differential branching fraction and angular analysis of the decay $B^{0} \to K^{*0} \mu^{+}\mu^{-}$,''
  JHEP {\bf 1308}, 131 (2013).


\bibitem{Aaij:2014bsa}
  R.~Aaij {\it et al.} [LHCb Collaboration],
  ``Measurement of $C\!P$ asymmetries in the decays $B^0 \rightarrow K^{*0} \mu^+ \mu^-$ and $B^+ \rightarrow K^{+} \mu^+ \mu^-$,''
  JHEP {\bf 1409}, 177 (2014).

\bibitem{Aaij:2014pli}
  R.~Aaij {\it et al.} [LHCb Collaboration],
  ``Differential branching fractions and isospin asymmetries of $B \to K^{(*)} \mu^+ \mu^-$ decays,''
  JHEP {\bf 1406}, 133 (2014).

\bibitem{ATLAS:2013ola}
  [ATLAS Collaboration],
  ``Angular Analysis of $B_{d} \to K^{\ast 0}\mu^{+}\mu^{-}$ with the ATLAS Experiment,''
  ATLAS-CONF-2013-038, ATLAS-COM-CONF-2013-043.


\bibitem{CMS:cwa}
  [CMS Collaboration],
  ``Angular analysis and branching ratio measurement of the decay $B^0$ to $K^{*0} \mu^+ \mu^-$,''
  CMS-PAS-BPH-11-009.


\bibitem{Faessler:2002ut}
  A.~Faessler, T.~Gutsche, M.~A.~Ivanov, J.~G.~Korner and V.~E.~Lyubovitskij,
  ``The Exclusive rare decays $B \to K(K^*) \bar{\ell} \ell$ and $B_c \to D(D^*) \bar{\ell} \ell$ in a relativistic quark model,''
  Eur.\ Phys.\ J.\ direct C {\bf 4}, 18 (2002).

\bibitem{Ebert:2010dv}
  D.~Ebert, R.~N.~Faustov and V.~O.~Galkin,
  ``Rare Semileptonic Decays of $B$ and $B_c$ Mesons in the Relativistic Quark Model,''
  Phys.\ Rev.\ D {\bf 82}, 034032 (2010).

\bibitem{Ball:1998kk}
  P.~Ball and V.~M.~Braun,
  ``Exclusive semileptonic and rare B meson decays in QCD,''
  Phys.\ Rev.\ D {\bf 58}, 094016 (1998).

\bibitem{Ball:2004rg}
  P.~Ball and R.~Zwicky,
  ``$B_{(d,s)} \to \rho, \omega, K^*, \phi$ decay form-factors from light-cone sum rules revisited,''
  Phys.\ Rev.\ D {\bf 71}, 014029 (2005).

\bibitem{Khodjamirian:2006st}
  A.~Khodjamirian, T.~Mannel and N.~Offen,
  ``Form-factors from light-cone sum rules with $B$-meson distribution amplitudes,''
  Phys.\ Rev.\ D {\bf 75}, 054013 (2007).

\bibitem{Khodjamirian:2010vf}
  A.~Khodjamirian, T.~Mannel, A.~A.~Pivovarov and Y.~M.~Wang,
  ``Charm-loop effect in $B \to K^{(*)} \ell^{+} \ell^{-}$ and $B\to K^*\gamma$,''
  JHEP {\bf 1009}, 089 (2010).

\bibitem{Ball:2005vx}
  P.~Ball and R.~Zwicky,
  ``SU(3) breaking of leading-twist $K$ and $K^*$ distribution amplitudes: A Reprise,''
  Phys.\ Lett.\ B {\bf 633}, 289 (2006).

\bibitem{Ali:1999mm}
  A.~Ali, P.~Ball, L.~T.~Handoko and G.~Hiller,
  ``A Comparative study of the decays $B \to (K, K^{*}) \ell^+ \ell^-$ in standard model and supersymmetric theories,''
  Phys.\ Rev.\ D {\bf 61}, 074024 (2000).

\bibitem{Becirevic:2006nm}
  D.~Becirevic, V.~Lubicz and F.~Mescia,
  ``An Estimate of the $B \to K^* \gamma$ form factor,''
  Nucl.\ Phys.\ B {\bf 769}, 31 (2007).


\bibitem{Liu:2011raa}
  Z.~Liu, S.~Meinel, A.~Hart, R.~R.~Horgan, E.~H.~Muller and M.~Wingate,
  ``A Lattice calculation of $B \to K^{(*)}$ form factors,''
  arXiv:1101.2726 [hep-ph].

\bibitem{Horgan:2013hoa}
  R.~R.~Horgan, Z.~Liu, S.~Meinel and M.~Wingate,
  ``Lattice QCD calculation of form factors describing the rare decays $B \to K^* \ell^+ \ell^-$ and $B_s \to \phi \ell^+ \ell^-$,''
  Phys.\ Rev.\ D {\bf 89}, 094501 (2014).


\bibitem{Bharucha:2010im}
  A.~Bharucha, T.~Feldmann and M.~Wick,
  ``Theoretical and Phenomenological Constraints on Form Factors for Radiative and Semi-Leptonic B-Meson Decays,''
  JHEP {\bf 1009}, 090 (2010).
  
\bibitem{Boyd:1997qw}
  C.~G.~Boyd and M.~J.~Savage,
  ``Analyticity, shapes of semileptonic form-factors, and anti-B $\to$ pi lepton anti-neutrino,''
  Phys.\ Rev.\ D {\bf 56}, 303 (1997).

\bibitem{Fu:2014pba}
  H.~B.~Fu, X.~G.~Wu, H.~Y.~Han and Y.~Ma,
  ``$B \to \rho$ transition form factors and the $\rho$-meson transverse leading-twist distribution amplitude,''
  J.\ Phys.\ G {\bf 42}, 055002 (2015).

\bibitem{Fu:2014cna}
  H.~B.~Fu, X.~G.~Wu, H.~Y.~Han, Y.~Ma and H.~Y.~Bi,
  ``The $\rho$-meson longitudinal leading-twist distribution amplitude,''
  Phys.\ Lett.\ B {\bf 738}, 228 (2014).

\bibitem{Choi:2007yu}
  H.~M.~Choi and C.~R.~Ji,
  ``Distribution amplitudes and decay constants for $(\pi, K, \rho, K^*)$ mesons in light-front quark model,''
  Phys.\ Rev.\ D {\bf 75}, 034019 (2007).

\bibitem{Ahmady:2014sva}
  M.~Ahmady, R.~Campbell, S.~Lord and R.~Sandapen,
  ``Predicting the $B \to K^*$ form factors in light-cone QCD,''
  Phys.\ Rev.\ D {\bf 89}, 074021 (2014).

\bibitem{Huang:2008zg}
  T.~Huang, Z.~H.~Li, X.~G.~Wu and F.~Zuo,
  ``Semileptonic $B(B_{(s)}, B_{(c)})$ decays in the light-cone QCD sum rules,''
  Int.\ J.\ Mod.\ Phys.\ A {\bf 23}, 3237 (2008).


\bibitem{Huang:1998gp}
  T.~Huang and Z.~H.~Li,
  ``$B \to K^* \gamma$ in the light cone QCD sum rule,''
  Phys.\ Rev.\ D {\bf 57}, 1993 (1998).

\bibitem{Huang:2001xb}
  T.~Huang, Z.~H.~Li and X.~Y.~Wu,
  ``Improved approach to the heavy to light form-factors in the light cone QCD sum rules,''
  Phys.\ Rev.\ D {\bf 63}, 094001 (2001).

\bibitem{Ball:2007zt}
  P.~Ball, V.~M.~Braun and A.~Lenz,
  ``Twist-4 distribution amplitudes of the $K^*$ and $\phi$ mesons in QCD,''
  JHEP {\bf 0708}, 090 (2007).

\bibitem{Agashe:2014kda}
  K.~A.~Olive {\it et al.} [Particle Data Group Collaboration],
  ``Review of Particle Physics,''
  Chin.\ Phys.\ C {\bf 38}, 090001 (2014).

\bibitem{Wu:2013ei}
  X.~G.~Wu, S.~J.~Brodsky and M.~Mojaza,
  ``The Renormalization Scale-Setting Problem in QCD,''
  Prog.\ Part.\ Nucl.\ Phys.\  {\bf 72}, 44 (2013).

\bibitem{Wu:2010zc}
  X.~G.~Wu and T.~Huang,
  ``An Implication on the Pion Distribution Amplitude from the Pion-Photon Transition Form Factor with the New BABAR Data,''
  Phys.\ Rev.\ D {\bf 82}, 034024 (2010).

\bibitem{BHL} S.J. Brodsky, T. Huang and G.P. Lepage, in Particles and Fields-2, edited by A.Z. Capri and A.N. Kamal (Plenum, New York, 1983), p.143; S.J. Brodsky, G.P. Lepage, T. Huang and P.B. MacKenzis, in Particles and Fieds 2, edited by A.Z. Capri and A.N. Kamal (Plenum, New York, 1983), p.83.

\bibitem{Cao:1997hw}
  F.~G.~Cao and T.~Huang,
  ``Large corrections to asymptotic $F(\eta_c \gamma)$ and $F(\eta_b \gamma)$ in the light cone perturbative QCD,''
  Phys.\ Rev.\ D {\bf 59}, 093004 (1999).

\bibitem{Ball:2006nr}
  P.~Ball and R.~Zwicky,
  ``$|V_{td} / V_{ts}|$ from $B \to V \gamma$,''
  JHEP {\bf 0604}, 046 (2006).

\bibitem{deTeramond:2012rt}
  G.~F.~de Teramond and S.~J.~Brodsky,
  ``Hadronic Form Factor Models and Spectroscopy Within the Gauge/Gravity Correspondence,''
  arXiv:1203.4025 [hep-ph].

\bibitem{Brodsky:2013npa}
  S.~J.~Brodsky, G.~F.~de T¨¦ramond and H.~G.~Dosch,
  ``Conformal Symmetry, Confinement, and Light-Front Holographic QCD,''
  Nuovo Cim.\ C {\bf 036}, 265 (2013).

\bibitem{Colangelo:2000dp}
  P.~Colangelo and A.~Khodjamirian,
  ``QCD sum rules, a modern perspective,''
  In *Shifman, M. (ed.): At the frontier of particle physics, vol. 3* 1495-1576.

\bibitem{Bourrely:2008za}
  C.~Bourrely, I.~Caprini and L.~Lellouch,
  ``Model-independent description of $B \to \pi l \nu$ decays and a determination of $|V_{ub}|$,''
  Phys.\ Rev.\ D {\bf 79}, 013008 (2009)
  [Phys.\ Rev.\ D {\bf 82}, 099902 (2010)].

\bibitem{Straub:2015ica}
  A.~Bharucha, D.~M.~Straub and R.~Zwicky,
  ``$B\to V \ell^+\ell^-$ in the Standard Model from Light-Cone Sum Rules,''
  arXiv:1503.05534 [hep-ph].

\bibitem{Aliev:1996hb}
  T.~M.~Aliev, A.~Ozpineci and M.~Savci,
  ``Rare $B \to K^* l^+ l^-$ decay in light cone QCD,''
  Phys.\ Rev.\ D {\bf 56}, 4260 (1997).

\bibitem{Beneke:2001at}
  M.~Beneke, T.~Feldmann and D.~Seidel,
  ``Systematic approach to exclusive $B \to V l^+ l^-, V \gamma$ decays,''
  Nucl.\ Phys.\ B {\bf 612}, 25 (2001).

\bibitem{Grinstein:2004vb}
  B.~Grinstein and D.~Pirjol,
  ``Exclusive rare $B\to K^* l^+ l^-$ decays at low recoil: Controlling the long-distance effects,''
  Phys.\ Rev.\ D {\bf 70}, 114005 (2004).

\bibitem{Akar:2014hda}
  S.~Akar [BaBar Collaboration],
  ``Penguin and rare decays in B{\small A}B{\small AR},''
  J.\ Phys.\ Conf.\ Ser.\  {\bf 556}, 012047 (2014).

\bibitem{Ali:2002jg}
  A.~Ali, E.~Lunghi, C.~Greub and G.~Hiller,
  ``Improved model independent analysis of semileptonic and radiative rare $B$ decays,''
  Phys.\ Rev.\ D {\bf 66}, 034002 (2002).

\bibitem{Descotes-Genon:2015hea}
  S.~Descotes-Genon and J.~Virto,
  ``Time dependence in $B\to V \ell\ell$ decays,''
  JHEP {\bf 1504}, 045 (2015).

\bibitem{Niehoff:2015qda}
  C.~Niehoff,
  ``On $B\to K^{(*)}\bar \nu\nu$ decays in and beyond the Standard Model,''
  PoS(EPS-HEP2015)553.

\bibitem{Ahmady:2015fha}
  M.~Ahmady, D.~Hatfield, S.~Lord and R.~Sandapen,
  ``The effect of $c\overline{c}$ resonances in the $B \to K^{*} \mu^+ \mu^-$ Branching Ratio and Forward-Backward Asymmetry,''
  arXiv:1508.02327 [hep-ph].

\bibitem{Sahoo:2015qha}
  S.~Sahoo and R.~Mohanta,
  ``Study of the rare semileptonic decays $B_d^0 \to K^* l^+ l^-$ in scalar leptoquark model,''
  arXiv:1507.02070 [hep-ph].


\bibitem{Descotes-Genon:2015xqa}
  S.~Descotes-Genon, L.~Hofer, J.~Matias and J.~Virto,
  ``Theoretical status of $B \to K^* \mu^+\mu^-$: The path towards New Physics,''
  J.\ Phys.\ Conf.\ Ser.\  {\bf 631}, no. 1, 012027 (2015).


\bibitem{Hofer:2015kka}
  L.~Hofer and J.~Matias,
  ``Exploiting the symmetries of $P$ and $S$ wave for $B\to K^*\mu^+\mu^-$,''
  JHEP {\bf 1509}, 104 (2015).


\bibitem{Descotes-Genon:2014joa}
  S.~Descotes-Genon, L.~Hofer, J.~Matias and J.~Virto,
  ``QCD uncertainties in the prediction of $B \to K^* \mu^+ \mu^-$ observables,''
  arXiv:1411.0922 [hep-ph].

\end{thebibliography}
\end{document}